\documentclass{scrartcl}
\usepackage[utf8]{inputenc}
\usepackage[T1]{fontenc}
\usepackage{amsmath,amssymb}
\usepackage{hyperref}
\usepackage{bm}
\usepackage{comment}
\usepackage{booktabs,array,tabularx}
\usepackage{xcolor}
\usepackage{graphicx}
\usepackage{tikz}
\usetikzlibrary{arrows.meta,patterns}
\usepackage{units}

\newcommand{\rm}[1]{\text{#1}}

\newtheorem{remark}{Remark}
\newtheorem{proposition}{Proposition}

\newcommand{\R}{\mathbb{R}}
\newcommand{\dd}{\,\mathrm{d}}
\newcommand{\vv}{\mathbf{v}}

\newcommand{\ww}{\mathbf{w}}
\newcommand{\nn}{\mathbf{n}}
\newcommand{\xx}{\mathbf{x}}
\newcommand{\yy}{\mathbf{y}}
\newcommand{\aaa}{\mathbf{a}}

\newcommand{\g}{\mathbf{g}}

\newcommand{\vphi}{\bm{\varphi}}

\newcommand{\Sref}{\hat{\mathcal S}}
\newcommand{\C}{\mathcal C}
\newcommand{\Div}{\operatorname{div}}
\newcommand{\ez}{\mathbf e_z}

\title{Elastic rebound in viscous fluid: an efficient rigid--elastic splitting with an asymptotic gap model}

\author{Thomas Richter\thanks{Otto von Guericke University Magdeburg, Germany, \url{https://numerics.ovgu.de}, \url{thomas.richter@ovgu.de}}
}

\begin{document}
\maketitle

\begin{abstract}
We formulate and test a model for an elastic body that falls under gravity
onto the bottom of a container filled with a viscous incompressible fluid and
rebounds without topological contact. The body is a rigid body plus a small,
quasi-static elastic deformation, represented by a fixed compliance operator.
In the near-contact patch, the thin fluid film is described by a Reynolds
equation with cavitation. This allows the use of coarse meshes that need not resolve the gap, enabling efficient simulation. 
The film is coupled implicitly to the elastic surface
displacement and to the rigid normal motion. The bulk fluid (ALE
Navier--Stokes) is coupled monolithically to the rigid body. For the
axisymmetric case with vertical translation, the coupled step reduces
\emph{exactly} to a small dense nonlinear problem for the film pressure, with
the bulk eliminated through a $2\times2$ affine map.

The model is implemented in NGSolve~\cite{ngsolve} and compared with the experiments of
\cite{HTR2021} and the Eulerian FSI simulations of~\cite{WRFH2021}. For a PTFE ball (St $\approx 50$) the approach agrees
with the Eulerian FSI to 1\%. The rebound height (maximal gap after contact) is 2.18\,mm, against
2.21\,mm measured experimentally and 2.06\,mm numerically for the Eulerian FSI, and a second small
bounce follows. For a
soft rubber ball (St $\approx 120$) the purely elastic model gives about
three times the measured rebound. Material damping is the most important
missing effect: a Kelvin--Voigt compliance with loss factor $\tan\delta=0.6$
gives 4.0\,mm, against 4.4\,mm measured. A planar version of the model is
compared with the recently published  2D benchmark of~\cite{CFFFKSTW2026}.

The radius of the film patch is the
only numerical parameter of the splitting. It is fixed by one a-priori rule
for all configurations, derived from a dry Hertz estimate of the approach,
and the results depend only weakly on it. 
\end{abstract}

\section{Introduction}\label{sec:introduction}

A smooth elastic body within a fluid-filled container is falling towards the bottom and bouncing off. This problem is easy to observe but its mathematical modeling and simulation is a hard problem. First of all, there is the no-collision result for the Navier--Stokes equations (e.g.~\cite{Hillairet2007,GVHW2015} among many others): if no-slip conditions act on body and bottom boundary, contact or release of contact is not possible. It must therefore be the elasticity of the solid body which explains the observation. An early hypothesis is a little fluid layer remaining in a gap between solid and boundary which transfers the forces, elastic energy being stored and released such that a rebound is possible without contact.

The main inspiration for the present work is the analysis
of~\cite{GSST2022}, where a reduced model, a system of ODEs for the gap
and an internal elastic degree of freedom, is used to investigate the possibility of contactless rebound. The drag in the reduced model is derived from the local shape of the body near the wall, $\psi(x)=\gamma|x|^{1+\alpha}$. The strength of its singularity as the gap closes is due to the geometry alone. This led~\cite{GSST2022} to conjecture that a
\emph{qualitative change in the shape of the solid} is necessary for a
physically meaningful rebound. Contactless rebound has since been reproduced
in fully resolved ALE simulations with geometric remeshing~\cite{Fara2024},
and it is the basis of the two-dimensional benchmark of~\cite{CFFFKSTW2026}.

The model of this paper is built on this insight. Section~\ref{sec:ehd} shows
by a scaling argument that the elastic flattening changes the drag law of the
film from $F\propto1/h$ for the sphere to $F\propto1/h^3$ for a flattened
surface, and that it does so only on a patch of radius $a\ll R$. We therefore
treat the body as rigid wherever its geometry has no influence on the drag,
and resolve the change of geometry only where it has.

The numerical simulation of this problem has long been hindered by a missing formulation that is able to deal with elasticity and contact at the same time. The standard approach for fluid solid interactions (rigid as well as elastic bodies) is the mapping of the problem to a reference domain in Arbitrary Lagrangian Eulerian (ALE) coordinates~\cite{Hirt1974,Richter2017}. While being highly accurate, robust and reliable, ALE simulations of the falling ball must resolve the gap. Considering realistic parameters this is mostly not feasible. However, recent improvement in hardware and ALE remeshing techniques makes this possible, although at high cost~\cite{Fara2024}. An alternative to ALE is the Fully Eulerian approach~\cite{Richter2013}, where the solid is mapped to the Eulerian system. The advantage is that this formulation is physically sound and that it does not need any artificial fields that might violate the numerical stability. Fully Eulerian formulations are able to model the elastic contact problem~\cite{Richter2013} and they are also able to cover physically relevant scenarios, see~\cite{Wahl2021} and in particular~\cite{Frei2024}. However, the computational cost, mostly arising in the demanding solution of highly nonlinear systems, prevents the practical application. 

An alternative to these two approaches that rely purely on fundamental modeling of fluid and elastic solid lies in reduced approaches that use specific contact models that do not need to resolve the little gap. They allow for larger time steps and mesh sizes and fast simulation. They do however usually need contact parameters that must be tuned to the specific situation. Different mechanisms are used to describe elastic contact, either based on penalty forces, on Lagrange multipliers~\cite{DinizdosSantos2008} or the Nitsche method~\cite{Nitsche1971,Chouly2013}. Recently there has been great advance in this direction and the contact zone is for instance modeled by means of an artificial thin fluid layer~\cite{Burman2020,Burman2022}.

This contribution aims at a model that is free of tuneable parameters and suitable for efficient implementation. It is based on an analysis of the relevant scales and coupling dynamics. We need to consider elasticity to get a rebound. However, elasticity acts on a local scale. A light plastic ball hitting the bottom at low velocity will not deform during the fall and even during contact, deformation -- although essential -- is minimal. We therefore propose a decoupled approach, where the dynamics is treated as a fluid rigid body interaction and the deformation is mostly taken as quasi-static linear elasticity. The size of the remaining gap and the elastic deformation is so small that the remaining fluid layer is mostly acting through its pressure. Hence, once close to contact we activate a patch, where the fluid is replaced by a film equation. This means that the minimal mesh size is not directly related to the width of the gap $h$ but to the size of the patch which has dimension $\sqrt{Rh}$, $R\gg h$ being the body's radius. 

Comparing to experimental data from our previous work~\cite{HTR2021} we will show that the model has predictive power and is able to give accurate values for the rebound height. The model works however in a narrow range of parameters. If the body is highly elastic, e.g. made from rubber, its deformation gets too large and our numerical results are off by more than a factor of 3 and model extensions are needed.

The following section gives an analysis of the relevant scales which is the basis for the mathematical model which is described in Section~\ref{sec:model}. Section~\ref{sec:disc} gives details on the discretization and the implementation and Section~\ref{sec:numerics} presents numerical test cases, in particular based on the experimental data from~\cite{HTR2021}. Finally, Section~\ref{sec:discussion} gives a discussion of the findings, limitations of the model and possible further extensions.

\section{Data and Scales}\label{sec:scales}

This section collects the experimental data we want to reproduce and uses
them to estimate the length and time scales of the problem. The estimates
rest on three classical pieces of mechanics that are not part of the
standard toolbox of numerical PDEs: lubrication theory for thin fluid films,
the surface response of an elastic half-space, and Hertz's theory of elastic
contact. Each is introduced in its simplest form, as far as it is needed. The
outcome is a strong separation of scales, which is the basis of the model in
Section~\ref{sec:model}. The notation is summarised in
table~\ref{tab:notation}.

\subsection{The experiments}\label{sec:experiments}

We aim at numerically reproducing the experiments of~\cite{HTR2021}.
Spherical balls of different materials and sizes (radius $R$) are released from rest in a
water--glycerin mixture inside a cylindrical container of diameter
$D=110$\,mm. They settle along the axis and impact on a massive
stainless-steel anvil that covers the bottom of the container. Each ball is held at its top by a vacuum tweezer,
well below the free surface, so that the top of the ball always starts at the
same height, $d_0+2R\approx167.8$\,mm above the anvil. The trajectories were
recorded at 1000 frames per second with a resolution of 8.89\,pixel/mm.

We consider only the cases in which the flow stays laminar. This excludes the
two steel balls and the large PTFE ball. The four remaining configurations
are listed in table~\ref{tab:material}. We refer to them by material and
diameter in millimetres: PTFE6, PTFE10, POM20 and Rubber22.

\begin{table}
\centering\footnotesize
\setlength{\tabcolsep}{3pt}
\begin{tabular}{llcccc}
\multicolumn{6}{l}{\emph{Fluid} (water--glycerin, 50\,\% by volume; identical for all cases)}\\
density & $\rho_f$ [kg\,m$^{-3}$] & \multicolumn{4}{c}{1141}\\
dynamic viscosity & $\eta$ [Pa\,s] & \multicolumn{4}{c}{0.008}\\
kinematic viscosity & $\eta/\rho_f$ [m$^2$\,s$^{-1}$] & \multicolumn{4}{c}{$7.0\cdot10^{-6}$}\\
surface tension & $\gamma$ [N\,m$^{-1}$] & \multicolumn{4}{c}{0.064}\\
container diameter & $D$ [mm] & \multicolumn{4}{c}{110}\\
gravity & $g$ [m\,s$^{-2}$] & \multicolumn{4}{c}{9.807}\\
\hline
\emph{Ball} & & \emph{PTFE6} & \emph{PTFE10} & \emph{POM20} & \emph{Rubber22}\\
diameter & $2R$ [mm] & 6 & 10 & 20 & 22\\
density & $\rho_s$ [kg\,m$^{-3}$] & 2122 & 2158 & 1351 & 1361\\
density ratio & $\rho_s/\rho_f$ & 1.86 & 1.89 & 1.18 & 1.19\\
mass & $m$ [g] & 0.24 & 1.13 & 5.66 & 7.59\\
Young's modulus & $E$ [MPa] & 670 & 670 & 2900$^{a)}$ & 2 (1.7--20.7)$^{b)}$\\
Poisson ratio & $\nu$ & 0.46 & 0.46 & 0.35$^{a)}$ & 0.4999\\
confinement & $2R/D$ & 0.055 & 0.091 & 0.18 & 0.20\\
\hline
\multicolumn{6}{l}{\emph{Experiment} (mean $\pm$ standard deviation over the runs)}\\
runs & & 3 & 10 & 3 & 2\\
initial gap & $d_0$ [mm] & $161.9\pm0.3$ & $157.7\pm0.7$ & $147.3\pm1.3$ & $146.1^{c)}$\\
terminal velocity & $v_T$ [m/s] & $0.332\pm0.003$ & $0.462\pm0.005$ & $0.272\pm0.005$ & $0.298\pm0.001$\\
term. Reynolds number & $\mathrm{Re}_T$ & 284 & 659 & 777 & 937\\
term. Stokes number & $\mathrm{St}_T$ & 59 & 139 & 102 & 124\\
impact velocity & $v_\mathrm{imp}$ [m/s] & $0.305\pm0.013$ & $0.435\pm0.009$ & $0.200\pm0.003$ & $0.278\pm0.007$\\
impact Stokes number & $\mathrm{St}_\mathrm{imp}$ & 54 & 130 & 75 & 116\\
restitution coefficient & $e=v_\mathrm{reb}/v_\mathrm{imp}$ & 0.45 & 0.63 & 0.68 & 0.31\\
first bounce height & $h_b$ [mm] & $2.31\pm0.14$ & $6.98\pm0.43$ & $7.29\pm0.45$ & $4.03\pm0.18$\\
Hertz contact time & $t_H$ [ms] & 0.11 & 0.17 & 0.20 & 3.5\\
\end{tabular}
\caption{Experimental configurations considered for the numerical study, taken from~\cite{HTR2021}. Diameters, densities and fluid data are from~\cite{HTR2021}, the value of $g$ and the elastic parameters of PTFE and rubber from~\cite{WRFH2021}. All kinematic quantities were extracted from the published trajectories (Mendeley Data, Hagemeier 2020).
$^{a)}$~Manufacturer data sheet of the POM balls (Kugel Pompel, $E=2.9$\,GPa, $\rho=1370$\,kg\,m$^{-3}$). The Poisson ratio is a typical value for POM (Goodfellow).
$^{b)}$~Range for HNBR; $E=2$\,MPa is used in the simulations.
$^{c)}$~Only one run; in the second run the ball was already moving in the first frame.}
\label{tab:material}
\end{table}

The quantities in table~\ref{tab:material} are obtained as follows. All
heights refer to the motion of the ball's centre and are measured from its
position at the first contact with the anvil. Hence $d_0$ is the initial
distance between the lowest point of the ball and the anvil, and $h_b$ is the
height of the apex after the first impact. The terminal velocity $v_T$ is the
maximal settling velocity (from a Savitzky--Golay derivative of the position
over 21\,ms). For PTFE6 and Rubber22 the velocity reaches a plateau. PTFE10
and POM20 reach a maximum and then decelerate noticeably while approaching the
wall, so that the velocity at impact is below $v_T$. The impact and rebound
velocities $v_\mathrm{imp}$ and $v_\mathrm{reb}$ are the slopes of linear fits to
the last four frames before and the first four frames after the impact. Their
ratio is the restitution coefficient $e$. The impact itself lasts less than one
frame and cannot be resolved at 1\,kHz; the contact time $t_H$ in the table
is the Hertz estimate~\eqref{eq:hertz} of Section~\ref{sec:hertz}.


\subsection{Notation and dimensionless numbers}\label{sec:notation}

Bold symbols denote vector and tensor fields, italic symbols scalars; the
subscripts $f$ and $s$ refer to the fluid and the solid. The vertical
coordinate $z$ points upwards, and the bottom of the container is $z=0$. The
centre of the ball is at height $Z(t)$ on the axis and moves with velocity
$V(t)=\dot Z(t)$, which is negative during the approach. We distinguish two
``gaps'':
$d(t)=Z(t)-R$ is the distance between the \emph{undeformed} sphere and
  the bottom. It is a single number and describes the rigid motion.
$h(r,t)$ is the actual local thickness of the fluid layer at distance
  $r$ from the axis. It includes the elastic deformation of the ball. 
Table~\ref{tab:notation} lists the symbols used throughout the paper.

\begin{table}
\centering\small
\begin{tabularx}{\textwidth}{>{$}l<{$}X}
\multicolumn{2}{l}{\emph{Geometry and motion}}\\
R & radius of the ball\\
Z(t),\ V(t)=\dot Z(t) & height of the centre above the bottom, vertical velocity\\
d(t)=Z(t)-R,\ d_0 & gap between the undeformed ball and the bottom; initial gap\\
h(r,t) & local thickness of the fluid film including the deformation\\
w(r,t) & vertical (upward) elastic displacement of the ball near the bottom\\
U & characteristic approach velocity; in the estimates $U=v_\mathrm{imp}$\\
a & radius of the ``pressure patch'' (region with large film pressure)\\
r_c & radius of the region in which the thin-film model is used, rule \eqref{eq:rc}\\
W,\ H & radius ($W=D/2$) and height of the container\\
\midrule
\multicolumn{2}{l}{\emph{Fluid}}\\
\rho_f,\ \eta & density, dynamic viscosity\\
\vv,\ p & velocity, (dynamic) pressure, see \eqref{eq:pdyn}\\
p_\mathrm{cav} & lower bound for the pressure (cavitation), Section~\ref{sec:nearphase}\\
\midrule
\multicolumn{2}{l}{\emph{Solid}}\\
\rho_s,\ m=\tfrac43\pi R^3\rho_s & density and mass of the ball\\
m_f=\tfrac43\pi R^3\rho_f & mass of the displaced fluid\\
E,\ \nu & Young's modulus, Poisson ratio\\
E^*=E/(1-\nu^2) & contact (plane-strain) modulus, Section~\ref{sec:ehd}\\
\mu_s=E/(2(1+\nu)),\ c_s=\sqrt{\mu_s/\rho_s} & shear modulus, speed of shear waves\\
\ww & elastic displacement field of the ball, Section~\ref{sec:kinematics}\\
\tau & retardation time of the viscoelastic damping, \eqref{eq:kv}\\
\midrule
\multicolumn{2}{l}{\emph{Dimensionless numbers}}\\
\mathrm{Re}=\rho_f U\,2R/\eta & Reynolds number\\
\mathrm{St}=\rho_s U\,2R/(9\eta) & Stokes number, \eqref{eq:St}\\
\varepsilon(h) & elasticity parameter, ratio of elastic deflection and gap, \eqref{eq:eps}\\
\end{tabularx}
\caption{Main notation. Subscripts $T$ and $\mathrm{imp}$ on Re and St indicate
that $U=v_T$ or $U=v_\mathrm{imp}$ is used.}
\label{tab:notation}
\end{table}

Besides the Reynolds number, the dimensionless number that controls whether a
ball bounces in a viscous fluid is the Stokes number
\begin{equation}\label{eq:St}
  \mathrm{St}=\frac{\rho_s U\,2R}{9\eta}
  =\frac1R\cdot\frac{mU}{6\pi\eta R}
  =\frac19\,\frac{\rho_s}{\rho_f}\,\mathrm{Re}.
\end{equation}
Since $6\pi\eta R U$ is the Stokes drag of a sphere, $mU/(6\pi\eta R)$ is the
distance a ball with initial speed $U$ would coast against this drag before
coming to rest; St is this stopping distance in units of $R$. Experiments show
that balls do not rebound for $\mathrm{St}\lesssim10$, and that the
restitution coefficient increases towards its value in air for large St
\cite{JZHR2001,GLP2002}. All four configurations have $\mathrm{St}\approx 50$--$130$ in the transitional regime.

Rigid balls do not bounce: for the Navier--Stokes equations with no-slip conditions, a rigid body cannot touch the wall in finite time~\cite{Hillairet2007,GVHW2015}.
A mechanism that stores energy is needed for a rebound, and the natural
candidate is elasticity. Elasticity alone is not enough, however.
\cite{GSST2022} show that the decisive effect is the change of
\emph{geometry} caused by the deformation: the film pressure flattens the
lower surface of the body, this changes the drag law qualitatively, and only
then can the stored elastic energy be released as a rebound while a fluid
layer of positive thickness remains and transfers the force throughout. We
make this precise by a scaling argument in Section~\ref{sec:ehd}. In the
lubrication literature the deformed shape is called the
\emph{elastohydrodynamic dimple}; it was computed for sphere--sphere
collisions by~\cite{DSH1986}.

\subsection{Thin films and the elastohydrodynamic scale}\label{sec:ehd}

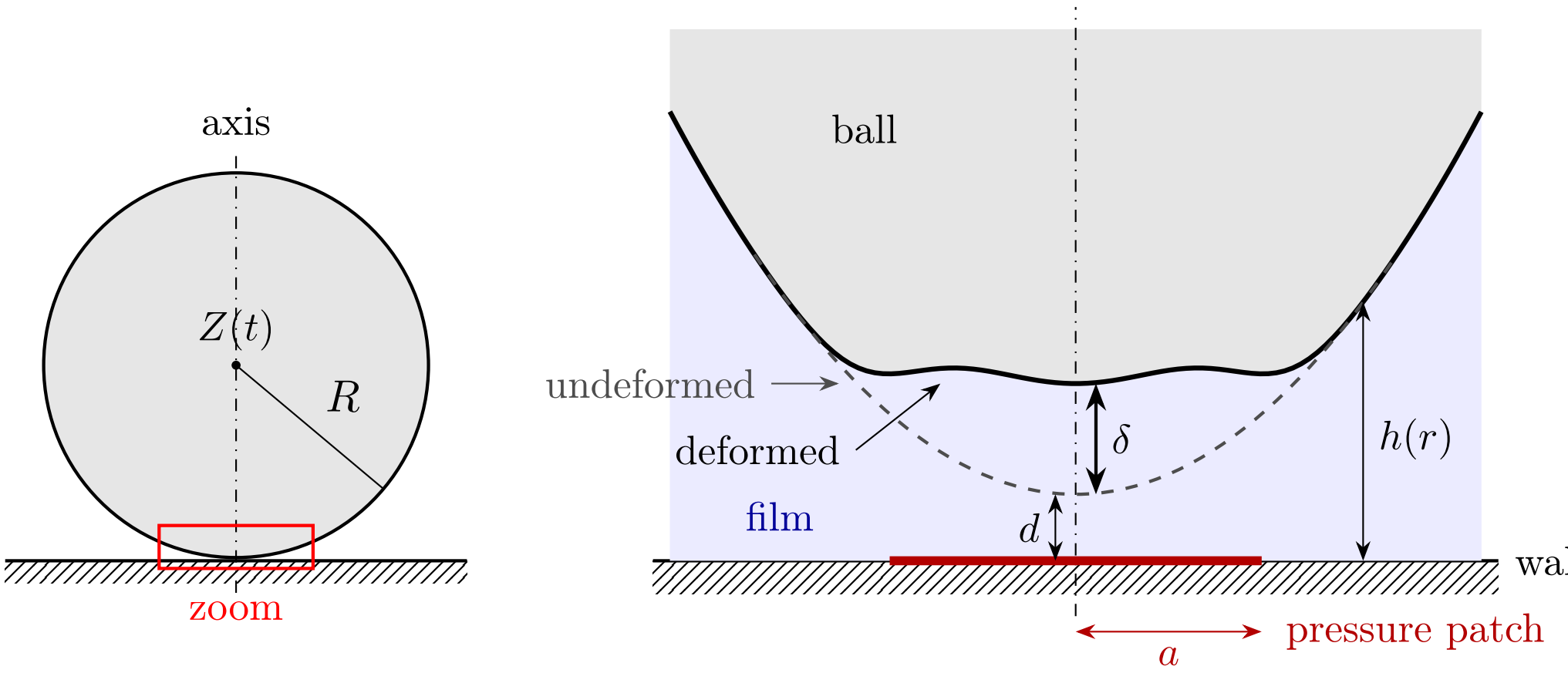
\begin{figure}
\centering
\begin{tikzpicture}[>=Stealth]
  \begin{scope}[scale=0.55]
    \fill[pattern=north east lines] (-3.6,-0.35) rectangle (3.6,0);
    \draw[thick] (-3.6,0) -- (3.6,0);
    \fill[gray!20] (0,3.05) circle (3);
    \draw[thick] (0,3.05) circle (3);
    \fill (0,3.05) circle (2pt);
    \draw (0,3.05) -- ({3*cos(-40)},{3.05+3*sin(-40)}) node[midway,above right] {$R$};
    \draw[dash dot] (0,-0.5) -- (0,6.4) node[above] {\small axis};
    \draw[red,thick] (-1.2,-0.12) rectangle (1.2,0.55);
    \node[red,below] at (0,-0.4) {\small zoom};
    \node at (0,3.6) {\small $Z(t)$};
  \end{scope}
  \begin{scope}[xshift=7.2cm,xscale=1.45,yscale=1.9]
    \fill[pattern=north east lines] (-2.5,-0.15) rectangle (2.5,0);
    \draw[thick] (-2.5,0) -- (2.5,0) node[right] {\small wall};
    \fill[blue!8,domain=-2.4:2.4,samples=100] (-2.4,0) -- plot (\x,{(0.3+0.3*\x*\x+0.5*exp(-((\x*\x)/1.2)*((\x*\x)/1.2)))}) -- (2.4,0) -- cycle;
    \fill[gray!20,domain=-2.4:2.4,samples=100] plot (\x,{(0.3+0.3*\x*\x+0.5*exp(-((\x*\x)/1.2)*((\x*\x)/1.2)))}) -- (2.4,2.4) -- (-2.4,2.4) -- cycle;
    \draw[very thick,domain=-2.4:2.4,samples=100] plot (\x,{(0.3+0.3*\x*\x+0.5*exp(-((\x*\x)/1.2)*((\x*\x)/1.2)))});
    \draw[dashed,thick,gray!60!black,domain=-1.9:1.9,samples=60] plot (\x,{0.3+0.3*\x*\x});
    \draw[dash dot] (0,-0.25) -- (0,2.5);
    \node at (-1.25,1.95) {\small ball};
    \node[blue!60!black] at (-1.75,0.2) {\small film};
    \draw[red!70!black,line width=2.2pt] (-1.1,0) -- (1.1,0);
    \draw[<->,red!70!black] (0,-0.32) -- (1.1,-0.32) node[midway,below] {\small $a$};
    \node[red!70!black,anchor=west] at (1.15,-0.32) {\small pressure patch};
    \draw[<->] (-0.12,0) -- (-0.12,0.3); \node[left] at (-0.12,0.15) {\small $d$};
    \draw[<->,thick] (0.12,0.3) -- (0.12,0.8); \node[right] at (0.12,0.55) {\small $\delta$};
    \draw[<->] (1.7,0) -- (1.7,{0.3+0.3*1.7*1.7+0.5*exp(-((1.7*1.7)/1.2)*((1.7*1.7)/1.2))});
    \node[right] at (1.7,0.55) {\small $h(r)$};
    \draw[gray!60!black,->] (-1.8,0.8) node[left] {\small undeformed} -- (-1.4,0.8);
    \draw[->] (-1.3,0.5) node[left] {\small deformed} -- (-0.8,0.8);
  \end{scope}
\end{tikzpicture}
\caption{Geometry of the near-contact region (right: zoom, heights strongly
exaggerated). $d=Z-R$ is the gap of the undeformed ball (dashed), $w$ the
upward elastic displacement of its surface, $\delta$ the deflection on the
axis, and $h(r)=d+r^2/(2R)+w(r)$ the local film thickness. The film pressure is
large only on the patch $r\lesssim a$, with $h\ll a\ll R$.}
\label{fig:ehdsketch}
\end{figure}

Near the bottom, the lower surface of the undeformed ball is
$z=Z-\sqrt{R^2-r^2}\approx d+r^2/(2R)$. The gap doubles at $r=\sqrt{2Rd}$,
and further out it widens such that the fluid can escape easily. The fluid is therefore squeezed in a
disc of radius
\[
  a\sim\sqrt{Rh},
\]
where $h$ is the typical film thickness in the disc (figure~\ref{fig:ehdsketch}).
For $h\ll R$ we have $h\ll a\ll R$: the region of high pressure is small
compared with the ball, but large compared to the film thickness.

In such a thin layer the flow is of lubrication type~\cite{Reynolds1886},
\cite[Ch.~4]{Batchelor1967}, \cite{HSJ2004}: the pressure is constant
across the film, and the \emph{radial volume flux per unit length of
circumference} is $J=-\frac{h^3}{12\eta}\,\partial_rp$. The factor $h^3$
makes thin films very hard to drain. Conservation of volume turns this into
the Reynolds equation, a two-dimensional (here: one-dimensional radial)
equation for the pressure, which replaces the Navier--Stokes equations in the
gap; it is derived as part of the model in Section~\ref{sec:nearphase}.

The ball pushes the volume $\pi a^2U$ per unit time out of the disc. It leaves
through the rim of length $2\pi a$, with flux $J\sim h^3p/(12\eta a)$ per unit
length. Equating both and using $a^2\sim Rh$ gives the pressure and force
\[
  p\sim\frac{\eta U a^2}{h^3}\sim\frac{\eta UR}{h^2},\qquad
  F\sim p\,a^2\sim\frac{\eta UR^2}{h}.
\]
For a rigid ball the Reynolds equation can be solved in closed form. With
$h=d+r^2/(2R)$, $\partial_th=V$ and $p\to0$ for $r\to\infty$,
\begin{equation}\label{eq:taylor}
  p(r)=-\frac{3\eta VR}{h(r)^2},\qquad
  F=\int_0^\infty p\,2\pi r\,\dd r=-\frac{6\pi\eta R^2}{d}\,V\qquad(d\ll R),
\end{equation}
which is the small-gap limit of the exact Stokes solution for a sphere
approaching a plane~\cite{Brenner1961,CB1967}. The force grows only like
$1/d$ and is proportional to $V$.

\textbf{Elastic deflection.}
Now let the ball be elastic. Since $a\ll R$, the ball looks, seen from the
pressure patch, like an elastic half-space loaded by the pressure $p$ on a
disc of radius $a$. The surface displacement of a half-space under a normal
point load was given by~\cite{Boussinesq1885}. For axisymmetric
loads the superposition was worked out by~\cite{Love1929}, see
\cite[Ch.~3]{Johnson1985} and \cite[\S8]{LL1986}. Linear elasticity on a
half-space has no intrinsic length scale, so the surface displacement must
scale like
\[
  \delta\sim\frac{p\,a}{E^*},\qquad E^*=\frac{E}{1-\nu^2}, 
\]
with an exact statement given in~\eqref{eq:boussinesq} (Section~\ref{sec:kinematics}).
For the deflection on the axis it gives $\delta=\frac{2}{E^*}\int_0^\infty p(s)\,\dd s$.
The \emph{contact modulus} $E^*$ is the combination of Young's modulus and
Poisson ratio that governs the normal surface response of a half-space: the
material below the load cannot expand freely sideways, which makes the surface
stiffer than a free bar by the factor $1/(1-\nu^2)$. 

The ratio of elastic deflection and gap follows by inserting the pressure
scale and $a\sim\sqrt{Rh}$,
\begin{equation}\label{eq:eps}
  \varepsilon(h):=\frac{\delta}{h}
  \sim\frac1h\cdot\frac{\eta UR}{h^2}\cdot\frac{\sqrt{Rh}}{E^*}
  =\frac{\eta\,U\,R^{3/2}}{E^*\,h^{5/2}} .
\end{equation}
The constant follows from the rigid squeeze film~\eqref{eq:taylor}: inserting
$p=3\eta UR/h(r)^2$ into the axis formula and using
$\int_0^\infty(1+t^2)^{-2}\dd t=\pi/4$ gives
\[
\frac{\delta}{d}=\frac{3\pi}{\sqrt2}\,\frac{\eta UR^{3/2}}{E^*d^{5/2}}\approx \frac{6.7\,\eta UR^{3/2}}{E^*d^{5/2}}.
\]
This is a first-order perturbation, valid as long as $\varepsilon\ll1$. Up to
a constant, \eqref{eq:eps} is the elasticity parameter of~\cite{DSH1986}. They show that the outcome of a collision is governed
by St and $\varepsilon$, and that a rebound requires St above a critical value
that depends on $\varepsilon$. The experiments of~\cite{BD1988} confirm this.
For $\varepsilon\ll1$ the ball behaves as a rigid body.
Elasticity changes the gap once $\varepsilon\sim1$, i.e.\ at
\begin{equation}\label{eq:hstar}
  h^*\sim R\left(\frac{\eta U}{E^* R}\right)^{2/5},\qquad
  a^*\sim\sqrt{R h^*}.
\end{equation}

Below $h^*$ the pressure flattens the surface. The pressure is largest on the
axis, so the centre is pushed back most, and the dimple forms: the film is
thinnest on a ring near the rim of the patch~\cite{DSH1986}. For the four
experiments, $h^*$ is between $0.7$ and $1.2\,\mu$m for the stiff balls and
$16\,\mu$m for the rubber ball (table~\ref{tab:scales}).

\textbf{The change of geometry changes the drag law.}
The scaling argument for $p$ and $F$ uses the shape of the body only through
the radius $a$ of the disc from which the fluid has to be squeezed out. For
the undeformed sphere, $a^2\sim Rh$ shrinks with the gap. Following
\cite{GSST2022}, we repeat the argument for a general local profile
$h(r)=d+\gamma r^{1+\alpha}$, where $\alpha=1$ is the sphere and larger
$\alpha$ means a flatter body. Then $a\sim(h/\gamma)^{1/(1+\alpha)}$, and
\begin{equation}\label{eq:dragalpha}
  F\sim p\,a^2\sim\frac{\eta U a^4}{h^3}
  \sim\eta U\,\gamma^{-\frac{4}{1+\alpha}}\,h^{\frac{1-3\alpha}{1+\alpha}} .
\end{equation}
This is the exponent that~\cite{GSST2022} derive rigorously from the
variational formulation of the Stokes problem in three dimensions. For the sphere, $\alpha=1$
gives the $1/h$ law~\eqref{eq:taylor}. The flatter the profile, the stronger
the singularity. In the limit $\alpha\to\infty$, a flat disc of fixed radius
$a$, one obtains the classical squeeze film between parallel
discs~\cite{Stefan1874}, $F=-3\pi\eta a^4V/(2h^3)$.

The elastic ball passes from one law to the other. For $h\gg h^*$ it is
effectively rigid and the drag follows the $1/h$ law~\eqref{eq:taylor}. Below
$h^*$ the film pressure flattens the surface. The radius of the high-pressure
region is then no longer $\sqrt{Rh}$ but is set by elasticity, $a\approx a^*$
at the onset and $a\approx a_H=\sqrt{R\delta_H}$ at maximal compression
(Section~\ref{sec:hertz}), and it no longer shrinks with the gap. The drag
law changes from $F\propto1/h$ to $F\propto1/h^3$. Compared with a rigid
sphere at the same gap, the force is larger by the factor
$a^4/(R^2h^2)\sim(\delta/h)^2$. With $\delta_H$ from table~\ref{tab:scales}
and the minimal film thickness of Section~\ref{sec:results}, this is about
$10^3$ at maximal compression.

A model of the rebound must therefore
contain two ingredients: elastic energy storage, and the change of geometry
in the film. Both act only on the patch $r\lesssim a$, with
$a/R\approx0.04$--$0.17$ (table~\ref{tab:scales}). Outside this patch the ball
is rigid to a very good approximation.

\subsection{Fast impacts: the Hertzian regime}\label{sec:hertz}

Whether the ball is stopped near $h^*$ depends on the Stokes number. For
small St the film absorbs the momentum of the ball before elasticity matters,
and there is little or no rebound. For large St the ball arrives at $h^*$ with
most of its momentum. The film then thins further, and its pressure rises
until it carries the full impact load. The deformation becomes much larger
than the gap, and the pressure distribution approaches that of a \emph{dry}
contact between an elastic sphere and a rigid plane. The thin film acts like
a lubricant that transmits the contact pressure. In elastohydrodynamic
lubrication of point contacts this is denoted as the approach of the film
pressure to the Hertz pressure~\cite[\S19]{HSJ2004}. Dry elastic contact is
described by Hertz's theory~\cite{Hertz1882}, which we summarise for a sphere
and a rigid plane following~\cite[Ch.~4]{Johnson1985} and~\cite[\S9]{LL1986}.

\textbf{Static Hertz contact.}
A sphere pressed against a rigid plane by a force $F$ is flattened on a disc
of radius $a$, and its centre approaches the plane by the indentation
$\delta$. The contact pressure has the shape of a half-ellipsoid,
\begin{equation}\label{eq:hertzstatic}
  p(r)=\frac{3F}{2\pi a^2}\sqrt{1-\frac{r^2}{a^2}},\qquad
  a^2=R\,\delta,\qquad
  F=\frac43\,E^*R^{1/2}\delta^{3/2}.
\end{equation}

\textbf{Hertzian impact.}
In an impact with velocity $U$, the kinetic energy $\frac12mU^2$ is converted
into this potential energy and back (quasi-static impact theory by~\cite{Hertz1882}; see~\cite[Ch.~11]{Johnson1985} and~\cite[\S9]{LL1986}). This gives the maximal indentation
$\delta_H$, the corresponding contact radius $a_H$ and, by integrating the
energy equation 
\[
\frac12m\dot\delta^2+\frac8{15}E^*R^{1/2}\delta^{5/2}
=\frac12mU^2,
\]
the duration of the contact,
\begin{equation}\label{eq:hertz}
  \delta_H=\Bigl(\frac{15\,mU^2}{16\,E^*\sqrt R}\Bigr)^{2/5},\qquad
  a_H=\sqrt{R\,\delta_H},\qquad
  t_H\approx 2.87\Bigl(\frac{m^2}{R\,E^{*2}\,U}\Bigr)^{1/5}.
\end{equation}
These formulas ignore the fluid and are used here only as scale estimates.
All four experiments are in this regime. The Hertzian indentation exceeds the
elastohydrodynamic gap scale by more than an order of magnitude,
$\delta_H/h^*\approx 15$--$22$ (table~\ref{tab:scales}). The simulations of
Section~\ref{sec:numerics} confirm these estimates up to the differences in
the impact velocity and the added mass of the fluid.

\begin{table}
\centering\footnotesize
\setlength{\tabcolsep}{3.5pt}
\begin{tabular}{lcccccccc}
case & $E$ [MPa] & St & $h^*$ [$\mu$m] & $\delta_H$ [$\mu$m] & $a_H$ [$\mu$m] ($a_H/R$) & $t_H$ [ms] & $R/c_s$ [$\mu$s] & $\dfrac{R/c_s}{t_H}$\\
\hline
PTFE6    & 670  & 54  & 0.74 & 11.5 & 186 (0.062)  & 0.11 & 9.1 & 0.08\\
PTFE10   & 670  & 130 & 1.16 & 25.7 & 358 (0.072)  & 0.17 & 15  & 0.09\\
POM20    & 2900 & 75  & 0.75 & 13.3 & 364 (0.036)  & 0.20 & 11  & 0.06\\
Rubber22 & 2    & 116 & 15.6 & 329  & 1903 (0.173) & 3.5  & 500 & 0.14\\
\bottomrule
\end{tabular}
\caption{Elastohydrodynamic and Hertz scales~\eqref{eq:hstar}, \eqref{eq:hertz}
of the four experiments. They are evaluated with the measured impact velocity
$U=v_\mathrm{imp}$, the contact modulus $E^*=E/(1-\nu^2)$ and the mass $m$ from
table~\ref{tab:material}, and $\eta=8$\,mPa\,s. Here $c_s=\sqrt{E/(2(1+\nu)\rho_s)}$ is the
shear-wave speed. The corresponding simulated values are reported in
Section~\ref{sec:results}.}
\label{tab:scales}
\end{table}

\subsection{Time scales}\label{sec:timescales}

Three time scales are involved: 1st, the fall, $d_0/v_T\approx0.5$\,s; 2nd, the contact, $t_H=0.1$--$3.5$\,ms; 3rd, the time an elastic wave needs to cross the ball, $R/c_s=9$--$500\,\mu$s, with the shear-wave speed $c_s$.
The ratio of the last two is $0.06$--$0.14$ (table~\ref{tab:scales}). During
the contact, elastic waves cross the ball many times, so the ball is at each
instant close to elastic equilibrium with the current load. In other words,
the elasticity is \emph{quasi-static}: the inertial term $\rho_s\ddot\ww$ in
the equations of elastodynamics can be neglected. The elastic energy can
\emph{not} be neglected, since it is what drives the rebound. 

For the soft
rubber ball the ratio is $0.14$ and neglecting inertia is questionable and we will numerically observe that the static elasticity model does not give a good prediction in the rubber case. 

The ratio of
the first two time scales, $1:10^{-2}$--$10^{-4}$, means that the numerical time step has
to be adapted by several orders of magnitude between fall and impact.

\subsection{Consequences: the design of the model}\label{sec:consequences}

The guiding principle, following~\cite{GSST2022}, is that the rebound is
caused by the change of geometry in the film~\eqref{eq:dragalpha}, together
with the storage of elastic energy. The model must therefore resolve both
exactly where they act, on the patch $r\lesssim a$, and may simplify
everything else. Together with the scale separation collected in this
section, this leads to the following model decisions:

\emph{A) Rigid motion plus small, quasi-static deformation.} The deformation
is localised and small, and the contact is slow compared with the elastic
waves (Section~\ref{sec:timescales}). The ball therefore moves as a rigid
body plus a small, quasi-static, linear elastic deformation: elastic inertia
is neglected, but the elastic energy, which drives the rebound, is kept. What
remains of the elasticity is a fixed linear \emph{compliance operator} that
maps surface loads to surface displacements.

\emph{B) The bulk sees a rigid ball.} The deformation is at most about $26\,\mu$m
for the stiff balls and $0.3$\,mm for the rubber ball. It is confined to
the near-contact patch. The bulk domain is therefore bounded by the
undeformed sphere and moves with the rigid motion only.

\emph{C) Thin film in the gap.} The film thickness is
$h/R\sim10^{-5}$--$10^{-4}$, the pressure patch $a/R\approx0.04$--$0.17$. Resolving this gap in a mesh cannot lead to a feasible simulation at reasonable computational cost. Close to the wall, the fluid in the gap is therefore
replaced by the Reynolds equation, a lower-dimensional model on a patch whose
resolution is set by $a$ and not by $h$. Hereby the dimension of the finite
element spaces is reduced by about 2 orders of magnitude. 

\emph{D) Half-space compliance.} Since $a\ll R$, the ball looks, seen from
the pressure patch, like an elastic half-space. The elastic displacement on
the patch is therefore computed from the film pressure with the Green's
function of the elastic half-space.

\emph{E) Cavitation.} During separation the film produces suction pressures
of similar magnitude as the compressive pressures during the approach, far
below what a liquid can sustain. The film pressure is therefore bounded
from below by a cavitation pressure $p_\mathrm{cav}$.

\emph{F) Axisymmetry.} The ball moves along the axis of the container
without rotation, and all computations are done in the meridian plane. This
restricts the model to small Reynolds numbers that allow for stable solutions
with cylindrical symmetry.

Two further consequences concern the numerical design rather than the model: 

\emph{G) Stiff film, added mass.} The film is stiff exactly in the gap, where elasticity plays a role. The pressure responds to the normal velocity of the surface with a stiffness of order $\eta a^2/h^3$, which is huge for small $h$. Explicit time splitting will be unstable and would require time steps comparable to the collision time. Film,
elasticity and rigid normal motion must be monolithically coupled. Moreover,
the bulk fluid and the ball have comparable densities
($\rho_s/\rho_f\approx 1$--$2$), so the added-mass effect is strong and the
coupling between rigid body and bulk fluid must be implicit as well
(Section~\ref{sec:splitting}).

\emph{H) Patch size.} The region in which the film model is used must contain
the Hertzian flattening: the rigid gap at its edge must stay large compared
with the indentation. The patch size is fixed by the same a-priori rule for
all configurations (Section~\ref{sec:patchradius}).

\section{Model}\label{sec:model}

In Section~\ref{sec:overview} we first state the full fluid--structure
interaction problem and recall the reductions derived in
Section~\ref{sec:consequences}. Section~\ref{sec:kinematics} treats the
elastic body, Section~\ref{sec:fluid} states the complete axisymmetric model
used for the simulations in this work and Section~\ref{sec:switch} gives the
rules for switching between its two configurations. Sections~\ref{sec:weak}
and~\ref{sec:energy} give the variational formulation and the energy balance.

\subsection{From the full problem to the reduced model}\label{sec:overview}

The ball occupies the stress-free reference configuration $\Sref\subset\R^3$
and is deformed by $\chi(\cdot,t):\Sref\to\R^3$. It occupies
$\mathcal B(t)=\chi(\Sref,t)$, and the fluid fills
$\mathcal F(t)=\Omega\setminus\overline{\mathcal B(t)}$ in the container
$\Omega$. The fluid is described in Eulerian, the solid in Lagrangian
coordinates $\yy\in\Sref$:
\begin{equation}\label{eq:full}
\begin{aligned}
  \rho_f\bigl(\partial_t\vv+(\vv\cdot\nabla)\vv\bigr)-\Div\sigma_f=\rho_f\g,
  \quad \Div\vv&=0,\quad \sigma_f=2\eta D\vv-p_\mathrm{abs}I &&\text{in }\mathcal F(t),\\
  \rho_s\,\partial_t^2\chi-\Div_{\yy}\mathbf P(\nabla_{\yy}\chi)&=\rho_s\g
  &&\text{in }\Sref,\\
  \vv\circ\chi=\partial_t\chi,\quad
  (\sigma_f\circ\chi)\,JF^{-T}\,\hat\nn&=\mathbf P\,\hat\nn
  &&\text{on }\partial\Sref ,
\end{aligned}
\end{equation}
where $F=\nabla_{\yy}\chi$ and $J=\det(F)$, and $\hat\nn$ is the outer normal of $\Sref$. Further, $D\vv=\frac12(\nabla\vv+\nabla\vv^T)$ is the rate of strain,
$p_\mathrm{abs}$ the absolute pressure, $\mathbf P$ the first Piola--Kirchhoff
stress of the elastic material. The interface conditions express continuity of
velocity (no slip) and of traction, see~\cite[\S3]{Richter2017}. On the walls of the container $\vv=0$; the
condition at the free surface (top) is specified in Section~\ref{sec:farphase}. No
contact condition is needed: the becomes extremely thin but it stays positive.

The reduced model  has two configurations (figure~\ref{fig:domain}): a
\emph{far phase}, in which a rigid ball moves in a Navier--Stokes fluid, and a
\emph{near-contact phase}, in which the film and the compliance are added.

\subsection{The elastic body: rigid motion plus small deformation}\label{sec:kinematics}

Let $\Sref$ have mass density $\rho_s$, total mass $m$ and centre of mass at
the origin. We write the deformation as a translation of a slightly
deformed body,
\begin{equation}\label{eq:floating}
  \chi(\yy,t)=\yy + \aaa(t)+\ww(\yy,t),\qquad \yy\in\Sref,
\end{equation}
with a translation $\aaa(t)\in\R^3$ and the elastic displacement $\ww$, measured in the frame that moves with the body (a ``floating frame'', see~\cite{CL1977} and \cite{Shabana2020}). Rotation is not considered as we assume an axisymmetric reduction, where the ball may only move in vertical direction and does not rotate. 

This splitting is not unique, since a small rigid
motion could be moved from $\aaa$ into $\ww$. It is fixed by the
Tisserand (mean-axis) constraints~\cite{CL1977}
\begin{equation}\label{eq:tisserand}
  \int_{\Sref}\rho_s\ww\dd\yy=0
\end{equation}
which say that the elastic displacement has no net translation (and no net
linearised rotation). By this constraint, $\aaa$ is the exact centre of
mass. Integrating the momentum balance over the body then gives the
Newton--Euler equations
$m\ddot\aaa=m\g+\mathbf F$, where $\mathbf F$ is the force exerted by the
fluid. 
For a ball released on the axis of a cylindrical container we have $\aaa=Z(t)\ez$. Only the vertical momentum balance remains:
\begin{equation}\label{eq:rigid1d}
  m\dot V = -(m-m_f)\,g + F_\mathrm{bulk}+F_\mathrm{film},\qquad \dot Z=V,
\end{equation}
with the displaced fluid mass $m_f$ and the forces of the bulk fluid and the
film defined in \eqref{eq:forces}. The buoyancy appears because the dynamic
pressure \eqref{eq:pdyn} is used throughout, including in the film.

\textbf{Quasi-static elasticity.}
We insert~\eqref{eq:floating} into the equations of elastodynamics, neglect
the elastic inertia $\rho_s\ddot\ww$ (Section~\ref{sec:timescales}) and
linearise in $\ww$. Then $\ww$ solves, on the \emph{fixed} reference body,
the static linear elasticity problem
\begin{equation}\label{eq:linel}
  -\Div\sigma_s(\ww)=\mathbf f_\mathrm{in}\ \ \text{in }\Sref,\quad
  \sigma_s(\ww)\,\hat\nn=\mathbf t\ \ \text{on }\partial\Sref,\quad
  \sigma_s(\ww)=2\mu_s D\ww+\lambda_s(\Div\ww)I,
\end{equation}
with the Lam\'e constants $\mu_s,\lambda_s$ and the fluid traction $\mathbf t$
expressed in the body frame. Because the frame is accelerated, an inertial
(d'Alembert) body force appears~\cite{Shabana2020},
$\mathbf f_\mathrm{in}=-\rho_s(\ddot\aaa-\g)$.
Problem
\eqref{eq:linel} is a pure traction (Neumann) problem. 
The solution is unique up to a rigid displacement, which is fixed
by~\eqref{eq:tisserand}. The solution map $\mathbf t\mapsto\ww$ is the
\emph{traction-to-displacement (compliance) operator} $\C$ of the free body.
It is linear, symmetric and positive semi-definite; $\frac12\langle\mathbf
t,\C\mathbf t\rangle$ is the stored elastic energy. $\C$ depends only on the
reference geometry and the material, never changes during the simulation, and
can be applied by back-substitution with a once-factorised stiffness matrix. 

\textbf{The patch displacement and half-space approximation.}
The elastic displacement enters the model only through the thin film and,
due to the anisotropy of the gap, only through the vertical displacement
of the surface on the contact patch
\[
  w:=\ez\cdot\ww\big|_{\Gamma_c}.
\]
Throughout, the italic scalar $w$ denotes this
component and the bold $\ww$ the full displacement field. The block
$\C_{cc}$ of $\C$ that maps the patch pressure to $w$ is dense and can be
precomputed.

Since $a\ll R$, the ball is approximated near the patch by the elastic
half-space bounded by the tangent plane at its lowest point. The vertical
surface displacement of a half-space under a normal surface pressure $p$ is
given by Boussinesq's Green's function~\cite{Boussinesq1885}, \cite[Ch.~3]{Johnson1985},
\[
  w(\xx)=\frac{1}{\pi E^*}\int_{\R^2}\frac{p(\bm\xi)}{|\xx-\bm\xi|}\dd\bm\xi,
  \qquad \xx\in\R^2 .
\]
For an axisymmetric pressure on the disc $r<r_c$, the angular
integration can be done explicitly~\cite{Love1929}, \cite[Ch.~3]{Johnson1985} and gives
\begin{equation}\label{eq:boussinesq}
  w(r)=\frac{4}{\pi E^*}\int_0^{r_c}\frac{s\,p(s)}{r+s}\,
  K\!\Bigl(\frac{4rs}{(r+s)^2}\Bigr)\dd s ,\qquad
  K(k^2)=\int_0^{\pi/2}\frac{\dd\theta}{\sqrt{1-k^2\sin^2\theta}},
\end{equation}
with the complete elliptic integral of the first kind $K$. 

\textbf{Viscoelastic damping.} Real materials dissipate part of the elastic
energy by internal friction. The simplest extension of the elastic compliance
is the Kelvin--Voigt solid, a spring and a dashpot in parallel, with stress
$E(\epsilon+\tau\dot\epsilon)$ for the strain $\epsilon$ and a retardation
time $\tau$. Applied to the patch displacement $w$ it reads
\begin{equation}\label{eq:kv}
  w+\tau\,\partial_t w=\C p .
\end{equation}
The purely elastic ball is the case $\tau=0$, which is the default in all
computations. Viscoelasticity gets relevant for Rubber22 and the choice of $\tau$ for the rubber ball is discussed in
Section~\ref{sec:kvsweep}.

\subsection{Mathematical model of the axisymmetric problem}\label{sec:fluid}

\begin{figure}[!tp]
\centering
\resizebox{\textwidth}{!}{%
\begin{tikzpicture}[
  font=\footnotesize,
  bc/.style={align=left, inner sep=1pt},
  wall/.style={line width=1.1pt},
  axisl/.style={dash dot, line width=0.6pt},
  sig/.style={line width=1.6pt, red!75!black},
  gc/.style={line width=1.6pt, blue!70!black},
  ballfill/.style={fill=gray!30, draw=black, line width=0.8pt},
  gfill/.style={fill=orange!25},
  >=latex,
  declare function={zrig(\x)=0.7+2.0*\x*\x;
                    zdef(\x)=0.7+2.0*\x*\x-0.6*exp(-2.5*\x*\x)+0.3*exp(-\x*\x/0.02);}]
\begin{scope}
  \def\W{2.8}\def\H{6.0}\def\R{0.75}\def\Zc{3.7}
  \fill[blue!4] (0,0) rectangle (\W,\H);
  \draw[ballfill] (0,\Zc-\R) arc[start angle=-90,end angle=90,radius=\R] -- cycle;
  \draw[wall] (0,0) -- (\W,0) -- (\W,\H);
  \draw[line width=0.8pt, densely dashed] (0,\H) -- (\W,\H);
  \draw[axisl] (0,0) -- (0,\Zc-\R); \draw[axisl] (0,\Zc+\R) -- (0,\H);
  \node at (0.38,\Zc) {$\mathcal B$};
  \node at (2.0,3.3) {$\mathcal F(t)$};
  \draw[->, thick] (1.0,\Zc+0.3) -- ++(0,-0.5) node[right, inner sep=1pt] {$V$};
  \draw[->, thick] (2.3,2.6) -- ++(0,-0.6) node[right] {$\mathbf g$};
  \draw[<->] (-0.25,0) -- node[left] {$Z$} (-0.25,\Zc);
  \draw[densely dotted] (-0.35,\Zc) -- (0,\Zc);
  \draw[<->] (1.0,0) -- node[right] {$d=Z-R$} (1.0,\Zc-\R);
  \draw[densely dotted] (0,\Zc-\R) -- (1.1,\Zc-\R);
  \node[bc, anchor=south] at (\W/2,\H+0.05) {$\Gamma_\mathrm{top}$: $v_z=0,\ \sigma_{rz}=0,\ \int p\,r=0$};
  \node[bc, anchor=west] at (\W+0.1,\H/2) {$\Gamma_\mathrm{side}$:\\ $\vv=0$};
  \node[bc, anchor=north] at (\W/2+0.4,-0.08) {$\Gamma_\mathrm{bot}$: $\vv=0$};
  \node[bc, rotate=90, anchor=south] at (-0.75,3.0) {$\Gamma_\mathrm{axis}$: $v_r=0,\ \partial_rv_z=0$};
  \draw[<-] ({\R*0.71},{\Zc+\R*0.71}) -- (0.95,\Zc+1.3) node[right, bc] {$\Gamma_{\mathcal B}$: $\vv=V\ez$};
  \node[anchor=south west, font=\small\bfseries] at (-1.0,\H+0.5) {(a) far phase, $d\ge d_\mathrm{act}$};
\end{scope}
\begin{scope}[shift={(4.55,0)}]
  \def\W{2.8}\def\H{6.0}\def\R{0.75}\def\Zc{0.97}\def\rc{0.36}
  \fill[blue!4] (0,0) rectangle (\W,\H);
  \fill[gfill] (0,0) -- (\rc,0) -- ({\rc},{\Zc-sqrt(\R*\R-\rc*\rc)}) arc[start angle={-90+asin(\rc/\R)}, end angle=-90, radius=\R] -- cycle;
  \draw[ballfill] (0,\Zc-\R) arc[start angle=-90,end angle=90,radius=\R] -- cycle;
  \draw[wall] (\rc,0) -- (\W,0) -- (\W,\H);
  \draw[gc] (0,0) -- (\rc,0);
  \draw[sig] (\rc,0) -- ({\rc},{\Zc-sqrt(\R*\R-\rc*\rc)});
  \draw[line width=0.8pt, densely dashed] (0,\H) -- (\W,\H);
  \draw[axisl] (0,\Zc+\R) -- (0,\H);
  \node at (0.35,\Zc+0.2) {$\mathcal B$};
  \node at (2.0,3.3) {$\mathcal F(t)$};
  \draw[->, thick] (2.3,4.6) -- ++(0,-0.6) node[right] {$\mathbf g$};
  \draw[gray] (-0.06,-0.06) rectangle (0.6,0.45);
  \node[gray, anchor=east] at (-0.06,0.25) {(c)};
  \node[bc, align=center, text width=2.6cm] at (\W/2-0.25,\H-0.8) {$\Gamma_\mathrm{top}$, $\Gamma_\mathrm{side}$, $\Gamma_\mathrm{axis}$:\\ as in (a)};
  \node[bc, anchor=north] at (\W/2+0.5,-0.08) {$\Gamma_\mathrm{bot}$: $\vv=0$};
  \draw[<-] ({\R*0.71},{\Zc+\R*0.71}) -- (0.95,\Zc+1.3) node[right, bc] {$\Gamma_{\mathcal B}$: $\vv=V\ez$};
  \draw[<-, red!75!black] (\rc+0.04,0.12) -- (1.0,0.45) node[anchor=south west, bc] {$\Sigma$:\\ $\sigma\nn=-p(r_c)\,\nn$};
  \draw[<-, blue!70!black] (0.2,-0.02) -- (0.2,-0.5) node[below, bc] {$\Gamma_c$: film};
  \node[anchor=south west, font=\small\bfseries] at (-0.3,\H+0.5) {(b) near-contact phase};
\end{scope}
\begin{scope}[shift={(8.75,0)}]
 \begin{scope}[yscale=1.35]
  \def\sx{3.4}
  \fill[gray!30] plot[domain=0:1.3, samples=80] ({\sx*\x},{zdef(\x)}) -- (1.3*\sx,4.3) -- (0,4.3) -- cycle;
  \fill[gfill] plot[domain=0:1, samples=60] ({\sx*\x},{zdef(\x)}) -- (\sx,0) -- (0,0) -- cycle;
  \fill[blue!4] (\sx,0) -- plot[domain=1:1.3, samples=30] ({\sx*\x},{zdef(\x)}) -- (1.3*\sx,0) -- cycle;
  \draw[line width=0.8pt] plot[domain=0:1.3, samples=100] ({\sx*\x},{zdef(\x)});
  \draw[densely dashed] plot[domain=0:1.0, samples=40] ({\sx*\x},{zrig(\x)});
  \draw[wall] (\sx,0) -- (1.3*\sx,0);
  \draw[gc] (0,0) -- (\sx,0);
  \draw[sig] (\sx,0) -- (\sx,{zrig(1)});
  \draw[axisl] (0,0) -- (0,4.3);
  \node at (0.8,2.9) {$\mathcal B$};
  \node at (1.15*\sx,0.4) {$\mathcal F(t)$};
  \node[orange!50!black] at (0.8*\sx,0.4) {$G(t)$};
  \draw[<->] (0.5*\sx+0.25,0) -- node[right, inner sep=1pt] {$h$} (0.5*\sx+0.25,{zdef(0.57)});
  \draw[|-|, thin] (0.5*\sx,{zdef(0.5)}) -- (0.5*\sx,{zrig(0.5)}) node[above left, inner sep=1pt] {$w$};
  \node[anchor=west, font=\scriptsize] at (0.3,3.95) {dashed: rigid surface};
  \node[anchor=north, blue!70!black] at (0.45*\sx,-0.04) {$\Gamma_c$};
  \node[anchor=north] at (1.17*\sx,-0.04) {$\Gamma_\mathrm{bot}$};
  \node[anchor=north] at (0,-0.04) {$r=0$};
  \node[anchor=north, red!75!black] at (\sx,-0.04) {$r_c$};
  \node[anchor=east, red!75!black] at (\sx-0.03,1.85) {$\Sigma$};
  \draw[->, thick, red!75!black] (1.27*\sx,1.0) -- node[above] {$Q$} (\sx+0.1,1.0);
 \end{scope}
  \node[anchor=south west, font=\small\bfseries] at (-0.3,6.5) {(c) zoom of (b)};
\end{scope}
\end{tikzpicture}}
\caption{Computational domain in the meridian half plane $\{r>0\}$ (not to
scale) with the boundary conditions. (a)~Far phase: Navier--Stokes in
$\mathcal F=\Omega\setminus\overline{\mathcal B}$. (b)~Near-contact phase:
the gap region $G$ (shaded) below the ball is removed from the bulk domain and
replaced by the Reynolds film~\eqref{eq:reynolds}--\eqref{eq:reyn_bc} on $\Gamma_c$; bulk and film are coupled through
the edge pressure $p(r_c)$ and the flux $Q$ across $\Sigma$. (c)~Enlarged
near-contact region of (b), gap and $w$ exaggerated. The film thickness $h$ includes the vertical
elastic surface displacement $w=\ez\cdot\ww$, which forms the dimple; the bulk domain is bounded by the
undeformed (rigid) ball surface.}
\label{fig:domain}
\end{figure}
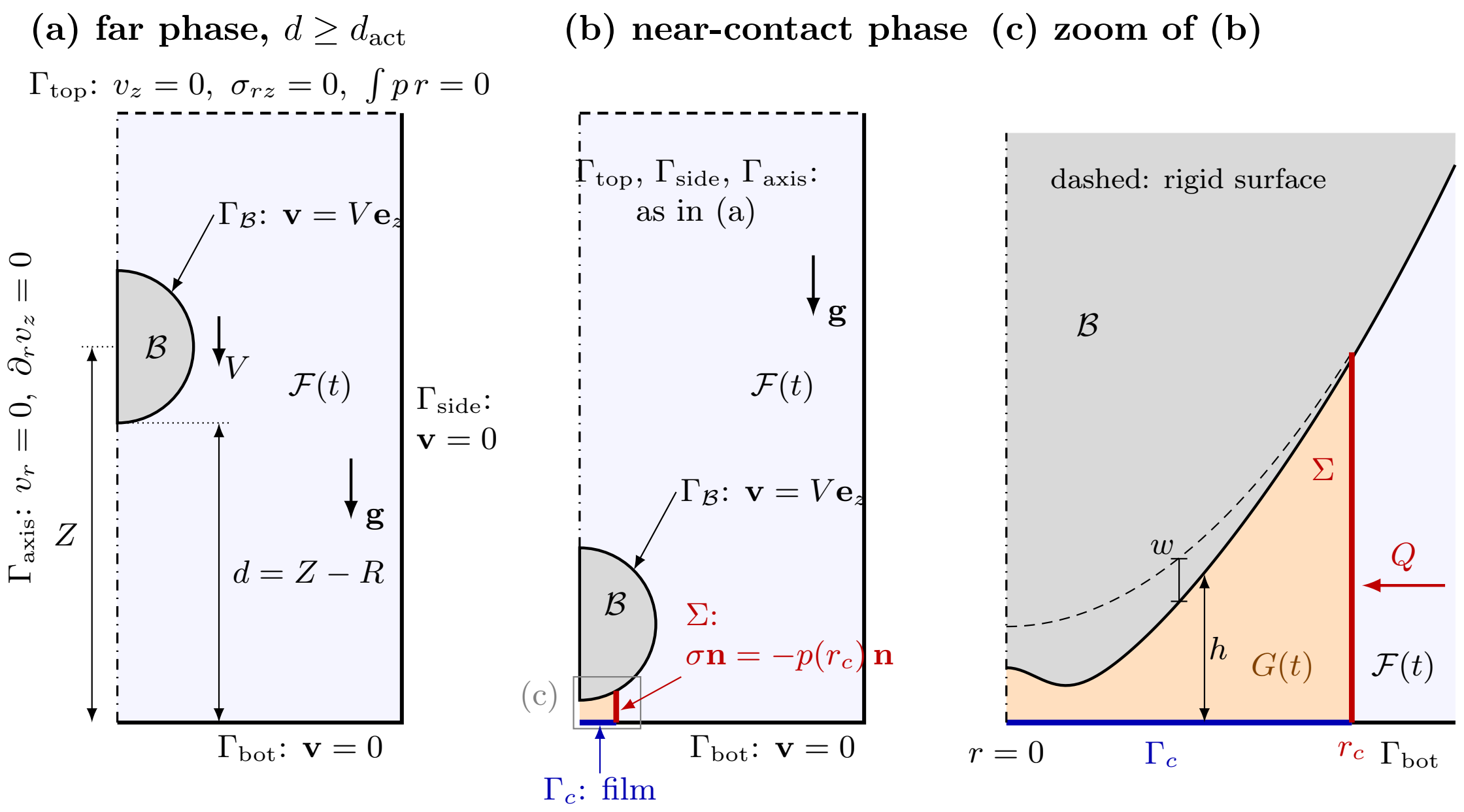

This section states the complete model that is solved in
Sections~\ref{sec:weak}--\ref{sec:results}: a rigid ball moving along the
axis of a cylindrical container, with the elastic deformation entering only
through the near-contact film. The model has two configurations
(figure~\ref{fig:domain}). In the \emph{far phase} the fluid fills the whole
container outside the ball. In the \emph{near-contact phase} the fluid in a
thin gap region $G$ below the ball is replaced by a Reynolds film on the patch
$\Gamma_c$, coupled to the compliance of the ball. The switching rules between
the two configurations (Section~\ref{sec:switch}) are part of the model.

\subsubsection{Geometry and unknowns}

We use cylindrical coordinates $(r,\theta,z)$ with the $z$-axis along the
axis of the container, pointing upwards. All fields are independent of
$\theta$, and there is no swirl, $v_\theta=0$. The container is the meridian
rectangle
\[
  \Omega=\{(r,z):\ 0<r<W,\ 0<z<H\},
\]
with the boundary parts
\[
  \Gamma_\mathrm{axis}=\{r=0\},\quad \Gamma_\mathrm{bot}=\{z=0\},\quad
  \Gamma_\mathrm{side}=\{r=W\},\quad \Gamma_\mathrm{top}=\{z=H\}.
\]
The ball of radius $R$ has its centre on the axis at height $Z(t)$:
\[
  \mathcal B(t)=\{(r,z):\ r^2+(z-Z(t))^2<R^2\},\qquad
  \Gamma_{\mathcal B}(t)=\partial\mathcal B(t),\qquad d(t)=Z(t)-R .
\]
$d$ is the distance between the undeformed ball and the bottom. The lower
surface of the ball is $z=Z-f(r)$ with $f(r)=\sqrt{R^2-r^2}$, $0\le r<R$.
Integrals over meridian sets carry the weight $r$, so they are quantities per
radian; the corresponding 3D quantities are $2\pi$ times larger.

The unknowns are the fluid velocity $\vv=v_r\mathbf e_r+v_z\ez$ and the
dynamic pressure $p$ in the bulk domain $\mathcal F(t)$, the position $Z(t)$
and velocity $V(t)=\dot Z(t)$ of the ball and, in the near-contact phase, the
film pressure $p(r,t)$ and the vertical elastic surface displacement
$w(r,t)=\ez\cdot\ww$ on $\Gamma_c$. The film pressure is denoted by the same letter as the bulk
pressure; the two are coupled by $p_\Sigma=p(r_c,t)$. The dynamic pressure is
the deviation from hydrostatics,
\begin{equation}\label{eq:pdyn}
  p_\mathrm{abs}=p_\mathrm{atm}+\rho_fg\,(H-z)+p ,
\end{equation}
so that gravity appears only as buoyancy in the equation of the ball.

\subsubsection{Bulk flow}

In $\mathcal F(t)$ the fluid satisfies the incompressible Navier--Stokes
equations, written in ALE form with respect to a mesh velocity $\vv_m$ that
equals $V\ez$ on $\Gamma_{\mathcal B}$ and has zero normal component on the
fixed boundaries:
\begin{equation}\label{eq:ns}
  \rho_f\bigl(\partial_t\vv|_{\hat x}+((\vv-\vv_m)\cdot\nabla)\vv\bigr)
  -\Div\sigma_f=0,\qquad \Div\vv=0,\qquad
  \sigma_f=2\eta D\vv-p\,I .
\end{equation}
In axisymmetric components, with the Eulerian time derivative
($\vv_m=0$),
\begin{equation}\label{eq:ns_comp}
\begin{aligned}
  \rho_f\bigl(\partial_tv_r+v_r\partial_rv_r+v_z\partial_zv_r\bigr)
   &=-\partial_rp+\eta\Bigl(\partial_r^2v_r+\frac1r\partial_rv_r-\frac{v_r}{r^2}+\partial_z^2v_r\Bigr),\\
  \rho_f\bigl(\partial_tv_z+v_r\partial_rv_z+v_z\partial_zv_z\bigr)
   &=-\partial_zp+\eta\Bigl(\partial_r^2v_z+\frac1r\partial_rv_z+\partial_z^2v_z\Bigr),\\
  \Div_r\vv:=\frac1r\partial_r(rv_r)+\partial_zv_z&=0 .
\end{aligned}
\end{equation}
The non-zero components of the rate of strain are
\[
  (D\vv)_{rr}=\partial_rv_r,\quad (D\vv)_{\theta\theta}=\frac{v_r}{r},\quad
  (D\vv)_{zz}=\partial_zv_z,\quad
  (D\vv)_{rz}=\tfrac12\bigl(\partial_zv_r+\partial_rv_z\bigr),
\]
so that $\sigma_{rz}=\eta(\partial_zv_r+\partial_rv_z)$ and
$\sigma_{rr}=-p+2\eta\,\partial_rv_r$.

\subsubsection{Motion of the ball}

With $m=\frac43\pi R^3\rho_s$ and the displaced fluid mass
$m_f=\frac43\pi R^3\rho_f$, the ball obeys \eqref{eq:rigid1d},
\begin{equation}\label{eq:ball}
  m\dot V=-(m-m_f)\,g+F_\mathrm{bulk}+F_\mathrm{film},\qquad \dot Z=V,
\end{equation}
\begin{equation}\label{eq:forces}
  F_\mathrm{bulk}=2\pi\!\int_{\Gamma_{\mathcal B}\cap\partial\mathcal F}\!
     \ez\cdot\sigma_f\,\nn_{\mathcal B}\;r\dd s,\qquad
  F_\mathrm{film}=\begin{cases}0&\text{far phase},\\[2pt]
     2\pi\displaystyle\int_0^{r_c}p\,r\dd r&\text{near-contact phase},\end{cases}
\end{equation}
where $\nn_{\mathcal B}$ is the outer unit normal of the ball. The buoyancy
$m_fg$ is that of the whole ball, including the part above $G$, because the
hydrostatic pressure acts on the whole surface and the film pressure is a
dynamic pressure as well. The film force uses the projected area; this is exact
for the vertical component of a pressure load, and the shear stress of the film
on the ball, of relative order $h/r_c$, is neglected.

\subsubsection{Far phase: Boundary and initial data}\label{sec:farphase}

The bulk domain is $\mathcal F(t)=\Omega\setminus\overline{\mathcal B(t)}$
(figure~\ref{fig:domain}a). The boundary conditions are
\begin{equation}\label{eq:bc_far}
\begin{aligned}
  &\Gamma_\mathrm{axis}: && v_r=0,\quad \sigma_{rz}=\eta\,\partial_rv_z=0 &&\text{(symmetry)},\\
  &\Gamma_\mathrm{bot},\ \Gamma_\mathrm{side}: && \vv=0 &&\text{(no slip)},\\
  &\Gamma_{\mathcal B}: && \vv=V\ez &&\text{(rigid ball, no slip)},\\
  &\Gamma_\mathrm{top}: && v_z=0,\quad \sigma_{rz}=0,\quad \int_{\Gamma_\mathrm{top}}p\,r\dd r=0
     &&\text{(flat free surface, ``slip'')}.
\end{aligned}
\end{equation}
The top models the free surface of the experiments as a flat, shear-free lid,
as in~\cite{WRFH2021}. The normalisation of $p$ makes the mean dynamic pressure
on the surface vanish, i.e.\ $p_\mathrm{abs}\approx p_\mathrm{atm}$ there by
\eqref{eq:pdyn}. This fixes the absolute pressure level, which matters for
cavitation. It is enforced by a Lagrange multiplier $\lambda$, which enters
the continuity equation as a uniform source on $\Gamma_\mathrm{top}$
(Section~\ref{sec:weak}). In the far phase the fluid volume is constant and
$\lambda=0$. In the near-contact phase,
$\lambda\,W^2/2=Q-r_c^2V/2$, which equals $\int_0^{r_c}\partial_tw\,r\dd r$
without cavitation: the multiplier takes up the elastic change of the gap volume,
which the rigid bulk geometry does not represent. 

The initial state is the ball at rest in fluid at rest:
\begin{equation}\label{eq:ic}
  Z(0)=R+d_0,\qquad V(0)=V_0=0,\qquad \vv(\cdot,0)=0 .
\end{equation}

\subsubsection{Near-contact phase: Boundary data}\label{sec:nearphase}

For a patch radius $r_c<R$ we define the gap region, the
film support and the interface (figure~\ref{fig:domain}b,c)
\begin{equation}\label{eq:G}
\begin{aligned}
  G(t)&=\{(r,z):\ 0<r<r_c,\ 0<z<Z(t)-f(r)\},\quad
  \Gamma_c=\{(r,0):\ 0<r<r_c\},\\
  \Sigma(t)&=\{(r_c,z):\ 0<z<Z(t)-f(r_c)\}.
\end{aligned}
\end{equation}
The bulk domain is $\mathcal F(t)=\Omega\setminus\overline{\mathcal B(t)\cup
G(t)}$. It is bounded by the undeformed ball: the elastic displacement acts
only in the film. On the parts of $\Gamma_\mathrm{axis}$, $\Gamma_\mathrm{bot}$,
$\Gamma_\mathrm{side}$, $\Gamma_\mathrm{top}$ and $\Gamma_{\mathcal B}$ that remain
on $\partial\mathcal F$, the conditions \eqref{eq:bc_far} hold. On $\Sigma$,
with the outer normal $\nn=-\mathbf e_r$ of $\mathcal F$,
\begin{equation}\label{eq:bc_sigma}
  \sigma_f\nn=-p_\Sigma\,\nn,\quad\text{i.e.}\quad
  \sigma_{rr}=-p_\Sigma,\ \ \sigma_{rz}=0,\qquad p_\Sigma(t)=p(r_c,t).
\end{equation}
The edge pressure is uniform over $\Sigma$, which requires
$h(r_c)\ll r_c$; see Section~\ref{sec:impl} for the consequences when this
fails.

On $\Gamma_c$ the film thickness is
\begin{equation}\label{eq:gap}
  h(r,t)=Z(t)-f(r)+w(r,t),
\end{equation}
where $w=\ez\cdot\ww$ is the vertical elastic displacement of the surface,
positive away from the bottom. It follows the film
pressure through the Kelvin--Voigt law~\eqref{eq:kv},
\begin{equation}\label{eq:kv_film}
  w+\tau\,\partial_tw=\C p,\qquad
  E^*=\frac{E}{1-\nu^2},
\end{equation}
with the Boussinesq operator $\C$ from \eqref{eq:boussinesq}.
Outside the patch all deformation is neglected.

\textbf{Reynolds equation.}
In $G$ the fluid layer is thin, $h\ll r_c$, and the Navier--Stokes equations
reduce to the lubrication approximation. We sketch the formal derivation; the
rigorous justification of the limit from Stokes to Reynolds is due to~\cite{BC1986}. Radial derivatives are smaller than vertical ones
by the factor $h/r_c$, and inertia is negligible: the reduced Reynolds number
$\rho_f U h^2/(\eta r_c)$ is below $10^{-2}$ in all cases. To leading order,
\[
  \partial_zp=0,\qquad \eta\,\partial_z^2v_r=\partial_rp,\qquad
  \frac1r\partial_r(rv_r)+\partial_zv_z=0\qquad\text{in }0<z<h(r,t).
\]
The pressure is constant across the film, $p=p(r,t)$. With no slip, $v_r=0$
at the bottom $z=0$ and on the ball surface $z=h$ (the ball moves only
vertically, and the tangential velocity of the elastic surface is neglected),
the radial velocity has the parabolic profile
$v_r=\frac{1}{2\eta}\partial_rp\,z(z-h)$. The radial volume flux per unit
length of circumference is
\[
  J:=\int_0^hv_r\dd z=-\frac{h^3}{12\eta}\,\partial_rp .
\]
Integrating the continuity equation across the film, with $v_z=0$ at $z=0$
and $v_z=\partial_th$ on the moving upper surface, gives the conservation of
volume, $\partial_th+\frac1r\partial_r(rJ)=0$, i.e.\ the Reynolds equation
\begin{equation}\label{eq:reynolds}
  \partial_t h-\frac1r\partial_r\Bigl(r\,\frac{h^3}{12\eta}\,\partial_r p\Bigr)=0
  \quad\text{on }(0,r_c).
\end{equation}
At each instant this is a nonlinear elliptic equation for $p$, which
degenerates as $h\to0$. Time enters only through $\partial_th$, which
contains the rigid velocity $V$ and the rate of deformation $\partial_tw$.
Together with $h=Z-f+w$ and $w=\C p$, \eqref{eq:reynolds} is a nonlocal
nonlinear problem for $p$; this is the \emph{elastohydrodynamic} problem.
The boundary conditions are
\begin{equation}\label{eq:reyn_bc}
  \lim_{r\to0}\ r\,h^3\partial_rp=0,\qquad
  \frac{r_c\,h^3}{12\eta}\,\partial_rp\Big|_{r=r_c}=Q(t),
\end{equation}
where
\begin{equation}\label{eq:flux}
  Q(t)=\int_\Sigma\vv\cdot\nn\,r\dd s=-r_c\,J(r_c)
\end{equation}
is the volume flux per radian from $\mathcal F$ into $G$ ($\nn$ points from
$\mathcal F$ into $G$). Bulk and film thus exchange two pieces of information
across $\Sigma$, in a coupling of Neumann--Neumann type: the bulk receives
the edge pressure of the film as a normal traction \eqref{eq:bc_sigma}, and
the film receives the volume flux of the bulk as a Neumann condition
\eqref{eq:reyn_bc}. The film problem alone determines $p$ only up to a
constant; the level is fixed through the bulk, whose pressure is normalised on
$\Gamma_\mathrm{top}$. Integrating \eqref{eq:reynolds} gives
$\frac{\dd}{\dd t}\int_0^{r_c}h\,r\dd r=Q$: the film conserves the fluid
volume exchanged with the bulk. With this coupling the interface terms cancel
in the energy balance (Section~\ref{sec:energy}).

\textbf{Cavitation.}
When the ball moves away from the wall, $h$ grows, and \eqref{eq:reynolds}
produces a suction pressure of the same magnitude as the compressive pressure
during the approach. Preliminary model runs (without cavitation) gave pressures down to $-190$\,bar and a qualitatively wrong rebound. Real liquids cannot sustain
such a tension. When the absolute pressure drops to the
vapour pressure, or below the saturation pressure of dissolved gas, bubbles
form and the pressure stays approximately constant. This is called
\emph{cavitation}. The simplest model (Swift--Stieber or half-Sommerfeld condition~\cite[chapter 10]{HSJ2004}) bounds the pressure from below
by $p_\mathrm{cav}$ and requires the Reynolds equation only where
$p>p_\mathrm{cav}$. Mathematically this is an obstacle problem, i.e.\ the
complementarity problem
\begin{equation}\label{eq:cav}
  p\ge p_\mathrm{cav},\qquad \mathcal R(p)\ge 0,\qquad
  (p-p_\mathrm{cav})\,\mathcal R(p)=0,
\end{equation}
where $\mathcal R(p)$ is the residual of the weak form of
\eqref{eq:reynolds}--\eqref{eq:reyn_bc}, tested with non-negative functions
(Section~\ref{sec:weak}). Where $p>p_\mathrm{cav}$, the Reynolds equation holds.
In the cavitated zone $p=p_\mathrm{cav}$, and the film volume is not conserved:
fluid (or vapour) is created there. 

By \eqref{eq:pdyn}, liquid at vapour pressure $p_\mathrm{vap}$ corresponds to
$p_\mathrm{cav}=p_\mathrm{vap}-p_\mathrm{atm}-\rho_fgH$. We always use 
$p_\mathrm{cav}=-1$\,bar, i.e.\ $p_\mathrm{abs}\approx0$; the vapour pressure of
water--glycerin (about $0.01$\,bar) and the hydrostatic load ($\rho_fgH=0.02$\,bar)
are neglected. 

This choice of $p_\mathrm{cav}=-1$\,bar is by physical reasoning. However, degassed liquids can sustain some tension, which shifts the limit towards no cavitation and gas-saturated liquids can release gas near
$p_\mathrm{abs}=p_\mathrm{atm}$, i.e.\ $p_\mathrm{cav}=0$ such that the exact value is uncertain. We will use $p_\mathrm{cav}=-1$\,bar in all simulations. 

\subsection{Switching between the phases}\label{sec:switch}

The near-contact phase is entered at the first time at which the ball
approaches and is closer than the activation gap,
\begin{equation}\label{eq:dact}
  d(t)<d_\mathrm{act}^\mathrm{eff}\quad\text{and}\quad V(t)<0,\qquad
  d_\mathrm{act}^\mathrm{eff}=\max\Bigl(d_\mathrm{act},\
  \Bigl(\frac{3\pi\,\eta\,|V|\,R^{3/2}}{\sqrt2\,E^*\,\varepsilon_\mathrm{act}}\Bigr)^{2/5}\Bigr).
\end{equation}
The second term is the gap at which the elasticity parameter of the rigid
squeeze film, 
\[
\varepsilon(h)=\frac{3\pi}{\sqrt2}\,\frac{\eta|V|R^{3/2}}{E^*h^{5/2}}=\frac{\delta}{h}
\]
(Section~\ref{sec:ehd}; cf.~\cite{DSH1986}), reaches $\varepsilon_\mathrm{act}$; it ensures that the
patch is active before elasticity matters (Section~\ref{sec:impl}). The far
phase is re-entered during separation,
\begin{equation}\label{eq:switch_off}
  d(t)>d_\mathrm{deact}^\mathrm{eff}\quad\text{and}\quad V(t)>0,\qquad
  d_\mathrm{deact}^\mathrm{eff}=\max\bigl(d_\mathrm{deact},\,0.75\,d_\mathrm{act}^\mathrm{eff}\bigr),
\end{equation}
with $d_\mathrm{act}^\mathrm{eff}$ taken at the preceding activation. Both criteria
use the rigid gap $d=Z-R$. We use $d_\mathrm{act}=0.01R$, $d_\mathrm{deact}=0.02R$
and $\varepsilon_\mathrm{act}=10^{-3}$, which gives $d_\mathrm{act}^\mathrm{
eff}=0.01R$ for PTFE6 and $0.048R$ for Rubber22. These are the only model parameters which could be called ``tuning parameters''. However, the sensitivity of the results to these values, tested in Section~\ref{sec:switchtest}, is very low.

At a switch, $Z$ and $V$ are continuous. At activation the film starts with
$p=0$ and $w=0$, which is consistent with the rigid ball of the far phase as
long as $\varepsilon\ll1$. At deactivation $p$ and $w$ are discarded. The
elastic energy left at that moment is about $10^{-7}$ (PTFE6) and
$4\cdot10^{-4}$ (Rubber22) of the kinetic energy at activation. The fluid velocity is
transferred to the new bulk domain and re-initialised by the projection
\eqref{eq:proj}, which keeps $V$ (Section~\ref{sec:impl}).

\subsection{Variational formulation in cylindrical coordinates}\label{sec:weak}

All integrals are written per radian, with weight $r$; $m'=m/(2\pi)$. The
rigid motion enters the velocity space through an extension field $E$, with
$E=1$ on the body and support in one element layer:
\[
  v_z=v_{z,0}+V E,\qquad \varphi_z=\varphi_{z,0}+B E,\qquad
  v_{z,0}=\varphi_{z,0}=0\ \text{on the body}.
\]
The fluid--body force then need not be computed explicitly; it is implicitly given by integration by parts. Find
$(v_r,v_{z,0},p,V,\lambda,p_\mathrm{film})$ such that, for all test functions
$(\varphi_r,\varphi_{z,0},\xi,B,\mu,\phi)$,
\begin{equation}\label{eq:weak}
\begin{aligned}
 &\bigl(\rho_f(\partial_t\vv+(\vv-\vv_m)\cdot\nabla\vv),\vphi\,r\bigr)
 +\bigl(2\eta\,D\vv:D\vphi\,r\bigr)-(p,\Div_r\vphi\,r)+(\Div_r\vv,\xi\,r)\\
&\quad  + p_\Sigma\!\int_\Sigma\vphi\cdot\nn\,r\dd s
-\lambda\int_{\Gamma_\mathrm{top}}\xi\,r\dd s+\mu\int_{\Gamma_\mathrm{top}}p\,r\dd s\\
 &\quad+\bigl(m'\dot V+(m'-m_f')g\bigr)B-B\int_0^{r_c}p\,r\dd r\\
 &\quad+\int_0^{r_c}\partial_t h\,\phi\,r\dd r
 +\int_0^{r_c}\frac{h^3}{12\eta}\partial_r p\,\partial_r\phi\,r\dd r
 -\phi(r_c)\int_\Sigma\vv\cdot\nn\,r\dd s=0,
\end{aligned}
\end{equation}
together with the complementarity condition \eqref{eq:cav}. Here
$\Div_r\vv=\partial_rv_r+v_r/r+\partial_zv_z$, and $D\vv:D\vphi$ contains the
 term $v_r\varphi_r/r^2$. The gap $h$ is given by \eqref{eq:gap}, so $h$
depends on $Z$ and, nonlocally, on $p$. The velocity spaces contain the
Dirichlet conditions of \eqref{eq:bc_far}: $v_r=0$ on $\Gamma_\mathrm{axis}$,
$\vv=0$ on $\Gamma_\mathrm{bot}$ and $\Gamma_\mathrm{side}$, $v_z=0$ on
$\Gamma_\mathrm{top}$; all other conditions are natural, including
$\partial_rv_z=0$ on the axis, $\sigma_{rz}=0$ on the top and
\eqref{eq:bc_sigma} on $\Sigma$. The multiplier $\lambda$ enforces the pressure
normalisation on $\Gamma_\mathrm{top}$ (Section~\ref{sec:farphase}). In the far
phase, the film unknowns and all terms on $\Sigma$ and $(0,r_c)$ are absent.
The stabilisation terms of Section~\ref{sec:impl} are not shown.

\subsection{Energy balance}\label{sec:energy}

Test \eqref{eq:weak} with $\vphi=\vv$, $\xi=p$, $B=V$, $\phi=p$. The coupling
terms on $\Sigma$ cancel. $\int\partial_th\,p\,r$ splits into the rigid part
$V\int p\,r$, which cancels with the work of the film on the body, and the
elastic part $\int\dot w\, p\,r=\frac{\dd}{\dd t}\mathcal E_\mathrm{el}$. For
symmetric $\C_{cc}$, $\mathcal E_\mathrm{el}=\frac12\int p\,w\,r$. One obtains
\begin{equation}\label{eq:energy}
\begin{aligned}
  &\frac{\dd}{\dd t}\Bigl[\tfrac12 m'V^2+(m'-m_f')gZ+\mathcal E_\mathrm{el}
  +\tfrac12\!\int_{\mathcal F}\rho_f|\vv|^2r\Bigr]
  +2\eta\|D\vv\|^2\\
  &\qquad +\int_0^{r_c}\!\frac{h^3}{12\eta}|\partial_rp|^2r
  +\tfrac12\!\int_{\Sigma\cup\Gamma_\mathrm{top}}\!\rho_f|\vv|^2(\vv-\vv_m)\cdot\nn\,r
  \le 0 .
\end{aligned}
\end{equation}
The inequality comes from cavitation. In the cavitated zone,
$(p-p_\mathrm{cav})\mathcal R=0$ with constant $p_\mathrm{cav}$.

The last term in \eqref{eq:energy} is the kinetic energy carried across the
open boundaries. On $\Sigma$ it is balanced, as the film has no kinetic energy.  Where fluid enters the bulk, i.e.\ $(\vv-\vv_m)\cdot\nn<0$,
it can feed energy into the bulk without bound. This is the classical
backflow instability of open boundaries~\cite{BF1994}. It is suppressed by
the backflow stabilisation described in Section~\ref{sec:impl}.

\section{Discretization}\label{sec:disc}

The discretization of this model is kept as simple as possible using low order methods in space and time. Also, simple direct solvers are used for all linear systems. The coupled model can be easily discretized using more sophisticated methods. In particular, higher-order methods can be used in both space and time. The linear systems are dominated by the Navier--Stokes equations in ALE coordinates, allowing for the use of fast iterative solvers. This is particularly necessary when extending the model to the three-dimensional case. 

For implementation we used the software framework NGSolve~\cite{ngsolve} for the bulk and NumPy~\cite{numpy} for dense matrices (the film equation). The code was developed with the help of \emph{Claude Code} based on the model \emph{Opus 5.5}. The complete implementation is published on Zenodo~\cite{implementation} along with a documentation of the code and a manual for reproducing the numerical examples. We therefore keep the following description short. 

\subsection{Time discretisation}\label{sec:splitting}

Time is discretised by the implicit Euler method with step size
$k_n=t_n-t_{n-1}$ using adaptive time-stepping based on the width of the gap and the performance of the nonlienar solver. Because of the added-mass effect and the stiffness of the
film, the bulk fluid, the rigid body and the film are coupled monolithically
in every step. Two approximations make the bulk part linear. First, the
convection term is linearised in Oseen form, with the convection field of the
previous step. Second, the geometry of the bulk domain is predicted
explicitly from the old state, and the bulk does not see the small elastic
deformation. The film problem stays fully nonlinear and implicit.

\textbf{The coupled system.}
Let the state $(Z^{n-1},V^{n-1},\vv^{n-1},\underline w^{n-1})$ at time
$t_{n-1}$ be given. The bulk domain is placed at the predicted position
\begin{equation}\label{eq:Zgeo}
  Z_\mathrm{geo}=Z^{n-1}+k_nV^{n-1},
\end{equation}
and the mesh velocity is $\vv_m=V^{n-1}\psi\,\ez$, with $\psi$ the harmonic
extension of $1$ on the body (Section~\ref{sec:impl}). On this domain,
$x=(v_r,v_{z,0},p,\lambda,V)\in\R^{N_b}$ collects the coefficients of the
bulk variables and of the rigid velocity in~\eqref{eq:weak}. The vectors
$\underline p,\underline w\in\R^{N_c}$ hold the nodal values of the film
pressure and of the elastic displacement on $(0,r_c)$, with node $N_c$ at
$r=r_c$. We use the following discrete operators.
\begin{itemize}
\item The Oseen matrix $A\in\R^{N_b\times N_b}$ of the bulk rows and of the
  $B$-row of~\eqref{eq:weak}. It contains
  $\partial_t\vv\approx(\vv-\vv^{n-1})/k_n$, the convection field
  $\vv^{n-1}-\vv_m$, the ball term $\frac{m'}{k_n}VB$ and the stabilisation
  terms (Section~\ref{sec:impl}). The right-hand side $b$ contains
  $\frac{\rho_f}{k_n}(\vv^{n-1},\vphi\,r)$ and
  $\bigl(\frac{m'}{k_n}V^{n-1}-(m'-m_f')g\bigr)B$.
\item The bulk coupling vector $s_i=\int_\Sigma\vphi_i\cdot\nn\,r\dd s$ and
  the unit vector $e_V$ of the rigid DOF, so that $Q=s^Tx$ and $V=e_V^Tx$.
\item The film coupling vectors $e_N=(0,\dots,0,1)^T$ and
  $m_i=\int_0^{r_c}\phi_i\,r\dd r$, so that $p_\Sigma=e_N^T\underline p$ and
  $F=m^T\underline p$.
\item The film mass matrix $M_f$, the Reynolds matrix
  $K(h)_{ij}=\int_0^{r_c}\frac{h^3}{12\eta}\,\phi_j'\phi_i'\,r\dd r$ and the
  compliance matrix $\C$ (Section~\ref{sec:impl}).
\end{itemize}
Implicit Euler for the Kelvin--Voigt law~\eqref{eq:kv_film} gives
$\underline w^n=\alpha\,\C\underline p+(1-\alpha)\underline w^{n-1}$ with
$\alpha=k_n/(k_n+\tau)$, so $\alpha=1$ in the static case $\tau=0$. The film
thickness at the nodes is
\begin{equation}\label{eq:hdisc}
  h^n=Z^{n-1}+k_nV^n-f+\underline w^n .
\end{equation}
The time derivative of the gap in the Reynolds equation is therefore
$(h^n-h^{n-1})/k_n=V^n+(\underline w^n-\underline w^{n-1})/k_n$. We define
the coupling matrix of rank two and the film matrix
\[
  B_\Sigma=s\,e_N^T-e_V\,m^T\in\R^{N_b\times N_c},\qquad
  S(h)=K(h)+\frac{\alpha}{k_n}\,M_f\C .
\]
In the near-contact phase, one time step is then the block system
\begin{equation}\label{eq:mono}
  \begin{pmatrix} A & B_\Sigma\\[2pt] -B_\Sigma^T & S(h^n)\end{pmatrix}
  \begin{pmatrix} x\\ \underline p\end{pmatrix}
  =\begin{pmatrix} b\\ \frac{\alpha}{k_n}M_f\underline w^{n-1}\end{pmatrix},
  \qquad h^n=h^n\bigl(e_V^Tx,\underline p\bigr)\ \text{by~\eqref{eq:hdisc}}.
\end{equation}
The system is linear in $x$ and nonlinear in $\underline p$. The conductance
$h^3$ depends on $\underline p$ through the elastic deformation, and on $x$
through $V^n$. With cavitation, the second block row is replaced by the
complementarity condition~\eqref{eq:cav} (see below). In the far phase there
are no film unknowns, and~\eqref{eq:mono} reduces to the linear system
$Ax=b$.

System~\eqref{eq:mono} shows where the Reynolds equation enters. The bulk
sees the film only through two scalars: the edge pressure $p_\Sigma$, which
is the normal traction on $\Sigma$ (column $s$), and the film force $F$ on
the ball (column $-e_V$). The film sees the bulk only through the flux $Q$,
which is the Neumann datum at $r_c$, and through the velocity $V^n$ in
$\partial_th$ and in $h^n$. The off-diagonal blocks are skew, so that
$x^TB_\Sigma\underline p-\underline p^TB_\Sigma^Tx=0$: the interface terms
cancel in the discrete energy balance, as they do in~\eqref{eq:energy}.

\textbf{Solution of one time step.}
Since $A$ does not depend on $\underline p$, the bulk unknowns can be
eliminated exactly. A time step consists of the following substeps.
\begin{enumerate}
\item \emph{Geometry.} Move the mesh to $Z_\mathrm{geo}$~\eqref{eq:Zgeo} and set the mesh
  velocity $\vv_m=V^{n-1}\psi\,\ez$. If the mesh displacement exceeds its
  limits, remesh first (Section~\ref{sec:impl}).
\item \emph{Bulk assembly.} Assemble $A$, $b$ and $s$ on the predicted
  geometry, and factorise $A$ once.
\item \emph{Elimination.} Three back-substitutions,
  $x_0=A^{-1}b$, $y_s=A^{-1}s$ and $y_V=A^{-1}e_V$, give the bulk solution as
  an affine function of the two coupling scalars,
  \[
    x=x_0-p_\Sigma\,y_s+F\,y_V,\qquad
    \begin{pmatrix}Q\\V\end{pmatrix}
    =\begin{pmatrix}s^Tx_0\\e_V^Tx_0\end{pmatrix}
    +M\begin{pmatrix}p_\Sigma\\F\end{pmatrix},
  \]
  with
  $M=\begin{pmatrix}-s^Ty_s&s^Ty_V\\-e_V^Ty_s&e_V^Ty_V\end{pmatrix}$.
  This is the Schur complement of $A$ in~\eqref{eq:mono}. The matrix
  $B_\Sigma^TA^{-1}B_\Sigma$ has rank two and is represented by the
  $2\times2$ matrix $M$.
\item \emph{Film.} Inserting the affine map into the second block row of
  \eqref{eq:mono} gives a problem for $\underline p$ alone,
  \[
    G(\underline p)=m\,V(\underline p)-e_N\,Q(\underline p)
    +S\bigl(h(\underline p)\bigr)\underline p
    -\frac{\alpha}{k_n}M_f\underline w^{n-1}=0 .
  \]
  This dense $N_c\times N_c$ system is solved by Newton's method with the
  Jacobian
  \[
    J=S(h)+\partial_h\bigl(K(h)\underline p\bigr)
      \bigl(\alpha\,\C+k_n\mathbf 1\,v^T\bigr)+m\,v^T-e_N\,q^T,
  \]
  where $v=\partial V/\partial\underline p=M_{21}e_N+M_{22}m$,
  $q=\partial Q/\partial\underline p=M_{11}e_N+M_{12}m$ and
  $\mathbf 1=(1,\dots,1)^T$. The bulk enters the Jacobian only through the
  rank-two term $m\,v^T-e_N q^T$ and through $v$ in $\partial h/\partial
  \underline p$. Without cavitation, a backtracking line search keeps $h>0$.
  With cavitation, $G_i=0$ is replaced by
  $\min\bigl(p_i-p_\mathrm{cav},\,G_i/d_i\bigr)=0$ with $d=|\mathrm{diag}\,J|$,
  and the system is solved by a primal--dual active-set method
  (Section~\ref{sec:impl}).
\item \emph{Recovery.} With the converged $\underline p$, compute $x$ from
  the affine map. This gives $\vv^n$ and $V^n=e_V^Tx$. Then set
  $Z^n=Z^{n-1}+k_nV^n$ and
  $\underline w^n=\alpha\,\C\underline p+(1-\alpha)\underline w^{n-1}$.
\item \emph{Control.} Choose the next step size, switch the patch if
  needed, and check the model assumptions (Section~\ref{sec:impl}).
\end{enumerate}
The first block row of~\eqref{eq:mono} is linear. Therefore the Newton
iteration of substep~4 produces exactly the iterates of Newton's method
applied to the full system~\eqref{eq:mono}: the elimination is not a
splitting. Each step costs one factorisation of $A$ and three
back-substitutions, independent of the number of Newton iterations, plus
dense $N_c\times N_c$ solves. Only the geometry is not treated fully
implicitly. The bulk domain is placed at $Z^{n-1}+k_nV^{n-1}$, while the film
thickness~\eqref{eq:hdisc} uses $Z^{n-1}+k_nV^n$. 

The extension to 3D with rotation is not trivial, since $B_\Sigma$ would
have rank $6+\dim(\text{traces on }\Sigma)$.

\subsection{Spatial discretization and solution}\label{sec:impl}

The implementation of the model is within the framework NGSolve~\cite{ngsolve} using NumPy~\cite{numpy} for all dense matrices (the film equation). The Navier--Stokes equations are discretized using the classical Taylor--Hood $P_2/P_1$ element on a curved mesh for resolving the ball. 
The rigid DOF $V$ enters via a simple discrete  extension $E_h$. Linear systems are solved using the direct solver UMFPACK~\cite{umfpack}. 

The film problem is discretized with $P_1$ elements on a graded 1D grid ($N_c=400$--$600$, resolution
$\approx 1\,\mu$m at $r=0$). The Jacobian is formed analytically, including the dense parts
$\C_{cc}$, $\partial V/\partial p$ and $\partial Q/\partial p$. A primal--dual active-set (semismooth Newton) method \cite{Hintermueller2010} is used for the
complementarity function $\min\bigl(p-p_\mathrm{cav},\,\mathcal R(p)/d\bigr)$.  The scaling $d=|\mathrm{diag}\,J|$ turned out to be essential: it gives both arguments the role of pressures. Without scaling, the active iteration stalled. 

\textbf{Stabilisation.} In the computations the backflow instability of
Section~\ref{sec:energy} appeared exactly when the film was cavitated: the
edge pressure no longer reacted to the flux, $\Sigma$ behaved like an open
boundary, and the fluid kinetic energy grew exponentially at fixed time step.
The remedy is the backflow stabilisation, also called directional do-nothing
condition~\cite{BraackMucha},
\begin{equation}\label{eq:backflow}
  -\tfrac12\rho_f\int_{\Sigma}\bigl((\vv^{n-1}-\vv_m)\cdot\nn\bigr)_-\,
  \vv\cdot\vphi\,r\dd s ,
  \qquad (x)_-=\min(x,0),
\end{equation}
which cancels the negative part of the flux in \eqref{eq:energy}. The same
term is used on an open top. We also add the consistent term
$\frac12\rho_f(\Div_r\vv^{n-1})\,\vv\cdot\vphi\,r$, which removes the energy
error caused by the convection field being only discretely
divergence-free.

\textbf{Compliance.} The Boussinesq kernel \eqref{eq:boussinesq} is computed using adaptive quadrature (log-singular) on elements near the
collocation node, and 8-point Gauss elsewhere. The complete elliptic integral is computing using \emph{ellipk} from the SciPy library~\cite{scipy}. The kernel is then assembled into the dense matrix $\C_{cc}$ whenever the patch is activated.

\textbf{Mesh motion and remeshing.} The mesh displacement is
$(Z_\mathrm{geo}-Z_\mathrm{ref})\psi\,\ez$, with $\psi$ being the harmonic $-\Delta \psi=0$ with  $\psi=1$ on the ball, $\psi=0$ on the walls, natural on the axis and on $\Sigma$, so that
$\Sigma$ stays at $r=r_c$. The mesh is regenerated (using OCC/Netgen meshing with local
refinement $h\approx \text{gap}/4$) when the displacement exceeds 35\% of the
thinnest bulk layer, of whenever  the ball moved by $1R$.

\textbf{Solution transfer.} The velocity is transferred between meshes by
interpolation.  To achive sufficient quality and avoid jumps of $V$ at remeshing it was essential to invert the ALE map, $\xx_\mathrm{ref}=\xx-\Delta Z\,\psi(\xx_\mathrm{ref})\,\ez$, by   fixed-point iteration \emph{to convergence}. 

\textbf{Topology switch.} Outside the patch phase the body is rigid, so
the patch must be switched on \emph{before} elasticity matters. Otherwise the
deformation that should already exist is created in a single step, as an
artificial jump of $w$. Activation is based on \eqref{eq:dact}. At activation, the mesh is regenerated without $G$, and the film starts with $p=0$, $w=0$.
The patch is deactivated during separation at the gap
\eqref{eq:switch_off}, and re-activated only while the ball approaches the
wall. 

\textbf{Consistent re-initialisation.} A switch between the two configurations changes the domain of the
bulk, and the transferred state is in general not compatible with the new
domain. In the implicit step this mismatch is projected out \emph{at once},
and the projection includes the rigid DOF. To protect the rigid velocity,
the fluid state is therefore re-initialised after every switch by the
projection
\begin{equation}\label{eq:proj}
  \vv^\mathrm{init}=\arg\min\bigl\{\|\vv-\vv^\mathrm{transf}\|_{\rho_f}:\ \Div_r\vv=0,\
  \vv=0\text{ on walls},\ \vv=V\ez\text{ on the body}\bigr\},
\end{equation}
computed as the limit $k\to0$ of the bulk step with the ball mass scaled up such that the  transferred state is nearly
consistent. The removed fluid kinetic energy is logged by the implementation and  negligible.

In the re-added fluid region, the transferred field is replaced before the
projection by the divergence-free rigid squeeze flow, given by the stream
function
\[
  \psi=\frac{V r^2}{2}s(\zeta),\quad \zeta=\frac zh,\quad s=3\zeta^2-2\zeta^3,\quad
  u_r=-\frac{V r}{2h}s'(\zeta),\quad
  u_z=Vs(\zeta)-\frac{Vr}{2}s'(\zeta)\zeta\frac{h'}{h}.
\]
For the transfer itself, the cutoff $\chi$ is the harmonic extension $\psi$ of
the new mesh, interpolated onto the old mesh, so it is the same smooth
function on both meshes. A geometric cutoff of width proportional to the gap
is not resolved near contact. At switch-off, the ALE extension on the old
mesh is continued into $G$ by its thin-layer form $z/h(r)$, so that the
inversion of the ALE map converges there.

\textbf{Time step control.} Outside the film phase we adapt the time step in step $n$ as 
\[
k_n=\min\bigl(k_{\max},\,1.3\,k_{n-1},\,0.1\,\text{gap}/|V|\bigr).
\]
In the film phase, $|V|$ is
replaced by the actual rate of change of the minimal film gap. During
separation the neck of the dimple stays thin while the ball already moves
fast, so $0.1\,h_{\min}/|V|$ would force steps of $10^{-4}T_0$ for thousands of
steps. In contact, two further limits apply: a cap of
$t_\mathrm{coll}/50$ (estimated from \eqref{eq:hstar}) while the collision is
fast, and a limit on the relative change of the film pressure (10\% of
$\max(|p|,\,0.05\,p_\mathrm{peak})$). Failed Newton steps are retried with $k/2$.
Checkpoints are written at every remesh, and a run can be restarted at the
contact phase.

\subsection{Choice of the patch radius}\label{sec:patchradius}

The patch radius is a numerical parameter of the splitting between kinetics and elastic deformation. It is no physical parameter, but we bound it from both sides by physical considerations in order to avoid the dependency on artificial tuning parameters. 
\begin{itemize}
\item \emph{Lower bound.} Outside the patch the bulk sees the undeformed
  body. While the centre of the ball approaches the wall by the indentation
  $\delta$, the rigid gap at the patch edge, $h_\mathrm{rig}(r_c)\approx
  d+r_c^2/(2R)$, must stay open. This requires $r_c^2/(2R)$ to be large
  compared with $\delta$.
\item \emph{Upper bound.} The film must be thin at its edge,
  $h(r_c)/r_c\approx r_c/(2R)\ll1$.
\end{itemize}
We use the same rule for all configurations, in 3D and in 2D:
\begin{equation}\label{eq:rc}
  \frac{r_c^2}{2R}=C_{rc}\,\delta^*,\qquad
  r_c=\min\bigl\{\max\{\sqrt{2RC_{rc}\delta^*},\,0.1R\},\,0.5R\bigr\},\qquad
  C_{rc}=4 .
\end{equation}
The radius is re-evaluated at every activation of the patch. $\delta^*$ is
a dry estimate of the approach for the current ball velocity $V$, computed
with the code's own compliance operator. For a family of contact radii $a$,
the Hertz pressure $p=p_0(a)\sqrt{1-r^2/a^2}$ is applied, with
$p_0=2aE^*/(\pi R)$ in the axisymmetric case and $p_0=aE^*/(2R)$ in the planar
case. The load is $P(a)=\int p$ (with the weights of the film force), and the
approach is $\delta(a)=(\C p)(0)$. $\delta^*$ then solves
$\int_0^{\delta^*}P\dd\delta=\frac12 m'V^2$. With the Boussinesq kernel this
reproduces $\delta_H$ of \eqref{eq:hertz} to $10^{-3}$. Defining the rule
through $\delta$ rather than through $a$ is what makes it transferable. In 3D,
$\delta=a^2/R$, and \eqref{eq:rc} is equivalent to $r_c=\sqrt{2C_{rc}}\,a\approx 2.8\,a$.
In 2D, the logarithmic kernel makes $\delta$ several times larger than $a^2/R$,
so a rule written in $a$ would give too small patches there. The value of
$C_{rc}$ is justified by the sensitivity test in Section~\ref{sec:rctest}:
for $C_{rc}=2$ the bulk layer at $\Sigma$ closes in both test cases, and for
$3\le C_{rc}\le8$ the results change only weakly. $C_{rc}=4$ keeps a margin
to the lower limit. The lower clip $0.1R$ only acts for slow
impacts (second bounces, the near-threshold case).

\section{Numerical Results}\label{sec:numerics}

\subsection{Experiments of \cite{HTR2021}}\label{sec:results}

We simulate the four configurations of
table~\ref{tab:material}. As in~\cite{WRFH2021}, the domain is a cylinder of
radius $0.055$\,m and height $0.2$\,m with a free-slip top. The ball starts at
rest at the measured mean initial gap $d_0$. Cavitation is limited at
$p_\mathrm{cav}=-1$\,bar. The patch radius follows from rule~\eqref{eq:rc}.
Each full simulation is continued at least until the second contact.

\textbf{Quantities.} Times are measured from $t_0$, when the centre of the
ball has fallen one radius~\cite{WRFH2021}. $v_T$ is the maximal settling
velocity and $t_\mathrm{cont}$ the time of the first contact. The simulated centre
position is evaluated in the same way as the camera data: it is sampled at
1\,kHz, and the impact and rebound velocities $v_\mathrm{imp}$, $v_\mathrm{reb}$ are
the slopes of line fits to the four frames before and after contact. The
restitution is $e=v_\mathrm{reb}/v_\mathrm{imp}$. $h_b$ is the height of the first
bounce, measured as the rise of the centre above its position at contact, and
$t_\mathrm{jump}$ is the time from contact to the apex. For short contacts, the
1\,kHz sampling underestimates the restitution: for PTFE6 it gives 0.50,
while the simulated velocities give 0.65. Since experiment and simulation are
evaluated identically, the values in table~\ref{tab:exp} are comparable.

\newcommand{\PaperRubberVT}{0.305}
\newcommand{\PaperRubberTcont}{0.531}
\newcommand{\PaperRubberVimp}{0.272}
\newcommand{\PaperRubberRestitution}{0.85}
\newcommand{\PaperRubberHb}{\ensuremath{14.5}}
\newcommand{\PaperRubberTjump}{\ensuremath{132}}
\newcommand{\PaperRubberSweepE}{0.81}
\newcommand{\PaperRubberSweepDjump}{\ensuremath{14.10}}
\newcommand{\PaperRubberSweepTjump}{\ensuremath{0.132}}

\begin{table}
\centering\footnotesize
\setlength{\tabcolsep}{3pt}
\resizebox{\textwidth}{!}{
\begin{tabular}{lcccccccc}
 & \multicolumn{2}{c}{PTFE6} & \multicolumn{2}{c}{PTFE10} & \multicolumn{2}{c}{POM20} & \multicolumn{2}{c}{Rubber22}\\
 & exp. (3) & model & exp. (10) & model & exp. (3) & model & exp. (2) & model\\
\hline
$v_T$ [m/s]          & $0.333\pm0.003$ & 0.315 & $0.462\pm0.005$ & 0.470 & $0.272\pm0.005$ & 0.285 & $0.298\pm0.001$ & 0.305\\
$t_\mathrm{cont}$ [s]   & $0.536\pm0.002$ & 0.559 & $0.388\pm0.002$ & 0.389 & $0.588\pm0.008$ & 0.572 & $0.544^{a)}$ & 0.531\\
$v_\mathrm{imp}$ [m/s]  & $0.305\pm0.013$ & 0.297 & $0.435\pm0.010$ & 0.451 & $0.200\pm0.003$ & 0.264 & $0.278\pm0.007$ & 0.271\\
$e$                  & $0.45\pm0.01$ & 0.50 & $0.63\pm0.02$ & 0.71 & $0.68\pm0.05$ & 0.71 & $0.31\pm0.01$ & 0.85\\
first-bounce height $h_b$ [mm]
                     & $2.31\pm0.14$ & 2.19 & $6.98\pm0.43$ & 7.85 & $7.29\pm0.45$ & 10.66 & $4.03\pm0.18$ & 14.41\\
$t_\mathrm{jump}$ [ms]  & $29.1\pm1.4$ & 28.8 & $54.1\pm2.3$ & 52.3 & $108\pm5$ & 113 & $88.7\pm0.6$ & 118\\
\hline
$r_c/R$ (first contact) & --- & 0.168 & --- & 0.201 & --- & 0.114 & --- & 0.492\\
$h_{\min}$ [$\mu$m]$^{b)}$ & --- & 0.28 & --- & 0.36 & --- & 0.27 & --- & 4.65\\
roughness [$\mu$m]~\cite{HTR2021} & 0.87 & --- & 0.87 & --- & 0.88 & --- & 3.84 & ---\\
\end{tabular}}
\caption{Simulation against all experimental runs (mean $\pm$ standard
deviation; number of runs in brackets). Purely elastic balls,
$p_\mathrm{cav}=-1$\,bar, corrected coupling on $\Sigma$, patch radius from
rule~\eqref{eq:rc}. The Rubber22 entries are generated from the full run.
$r_c/R$ is the a-priori model patch size, $h_{\min}$ is a simulation result,
and roughness is the measured experimental value; --- marks not applicable.
$^{a)}$~One run only (the second run has no resting phase).
$^{b)}$~Minimal film thickness during the first contact.}
\label{tab:exp}
\end{table}

\begin{figure}
\centering
\includegraphics[width=\textwidth]{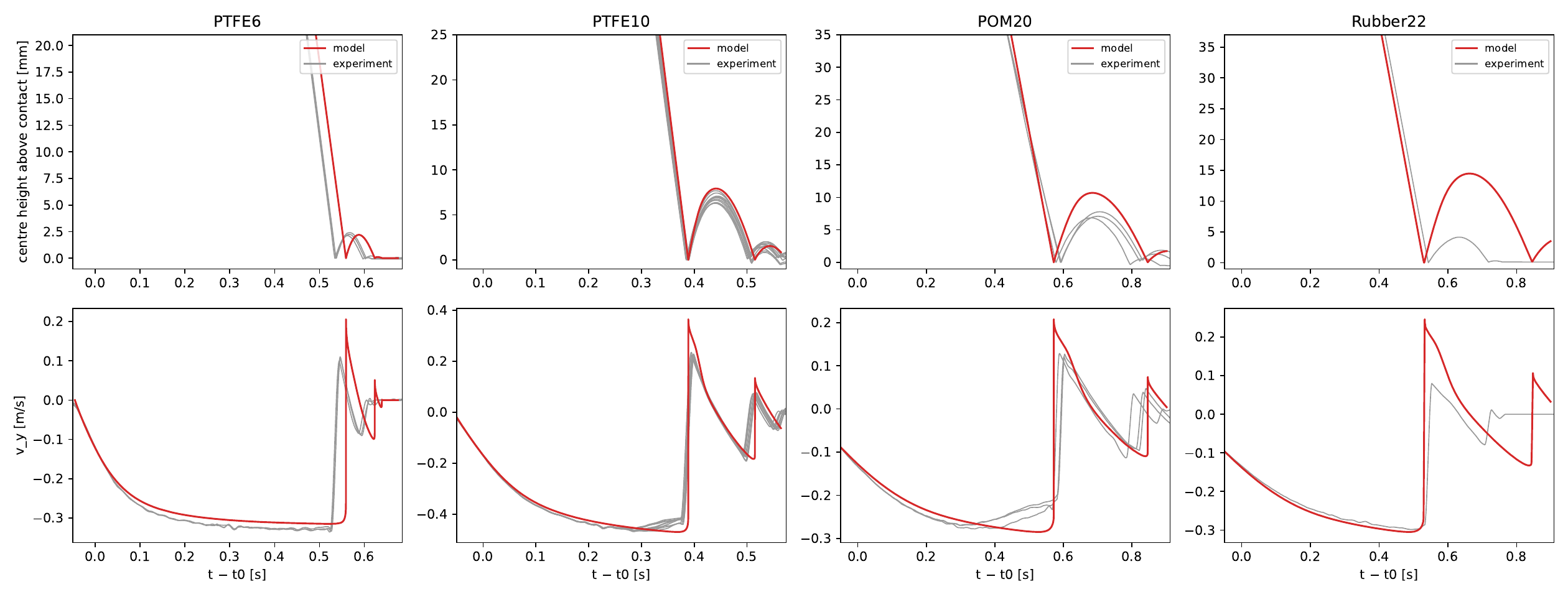}
\caption{Simulation (red) against all experimental runs (grey), times
relative to $t_0$. Top: height of the centre above its position at the first
contact. Bottom: vertical velocity. For Rubber22 only the run with a resting
phase is shown.}
\label{fig:compare_all}
\end{figure}

\begin{figure}
\centering
\includegraphics[width=\textwidth]{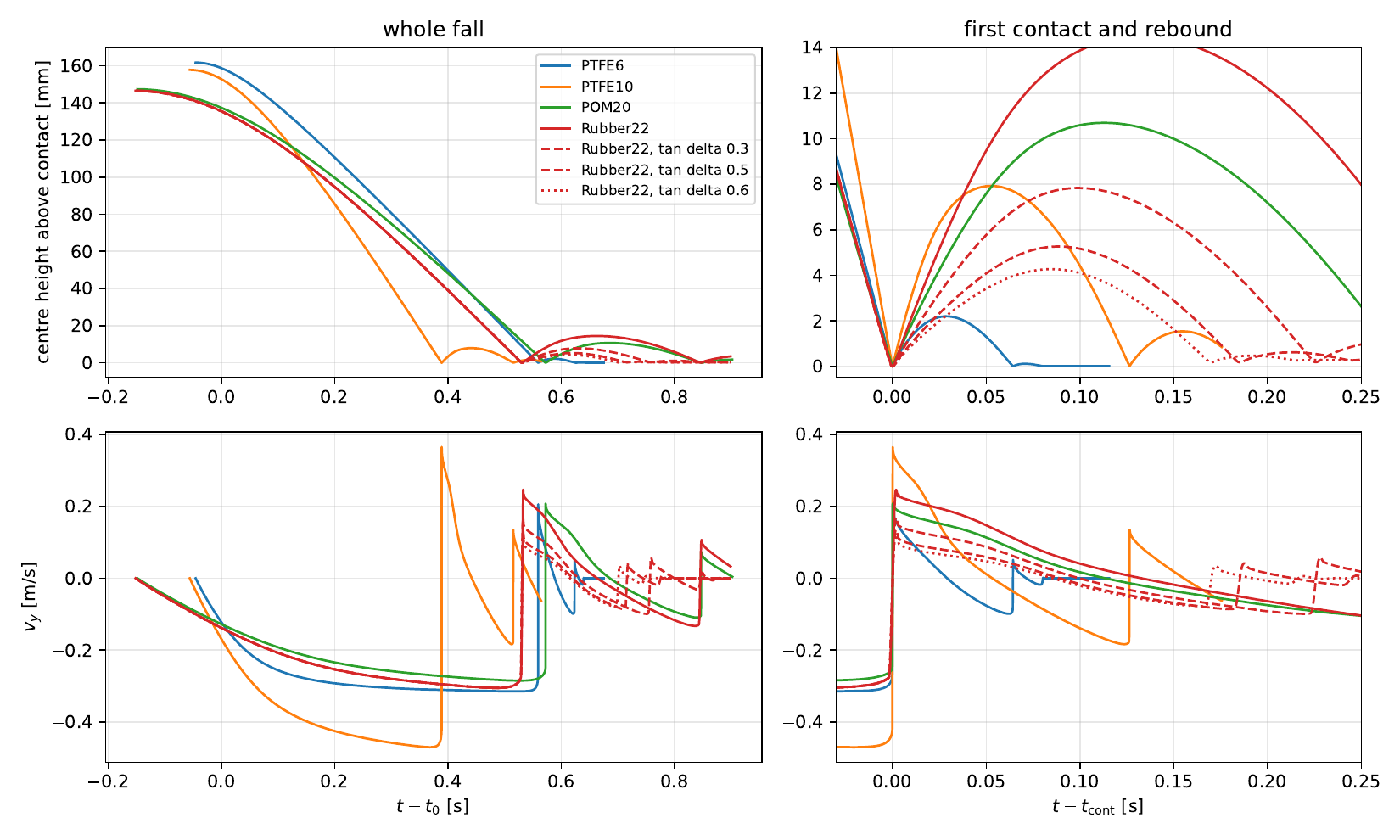}
\caption{All simulations. Top: height of the centre above its position at
the first contact. Bottom: vertical velocity. Left: the whole fall, times
relative to $t_0$. Right: first contact and rebound, times relative to
$t_\mathrm{cont}$. Rubber22 is shown purely elastic and with Kelvin--Voigt
damping (Section~\ref{sec:kvsweep}).}
\label{fig:all_sims}
\end{figure}

Figure~\ref{fig:all_sims} collects all simulated trajectories.

\textbf{PTFE6.} The settling velocity is 5\% below the experiment (0.315
against 0.333\,m/s), so the ball arrives 23\,ms late. The Eulerian FSI
of~\cite{WRFH2021} shows the same deficit (table~\ref{tab:fsi}); both
simulations agree with each other to 1\% before contact. The deficit is
therefore not caused by the splitting. Instead we attribute it to the uncertainty in the ball's release mechanism and the modeling of the surface with fixed geometry. Height and duration of the bounce agree
with the experiment within the scatter of the runs (2.19 against
$2.31\pm0.14$\,mm, 28.8 against $29.1\pm1.4$\,ms). The ball comes back and
performs a second, small bounce of $\approx0.11$\,mm, as observed in the
experiments. The maximal indentation is $10.6\,\mu$m and the radius of the
pressure patch $\approx150\,\mu$m, which agrees with the dry Hertz estimates
$\delta_H=11.5\,\mu$m and $a_H=186\,\mu$m of table~\ref{tab:scales} up to the
added mass of the fluid and the difference in the impact velocity.

\textbf{PTFE10.} The approach is reproduced very well. The settling
velocity agrees to 1.7\%, and the time of contact to 1\,ms. Like the
experiment, the model shows the deceleration in front of the wall, although
it is weaker (figure~\ref{fig:compare_all}). The bounce is 12\% too high (7.85
against $6.98\pm0.43$\,mm), and the restitution 0.71 against 0.63. Its duration
is correct. The maximal indentation is $27.2\,\mu$m (Hertz estimate:
$25.7\,\mu$m, table~\ref{tab:scales}).

\textbf{POM20.} Up to $t-t_0\approx0.45$\,s, model and experiment agree.
After that, the measured ball decelerates almost linearly until the impact,
which Hagemeier et al.\ attribute to a rotation of the ball by about
$180^\circ$ during settling~\cite{HTR2021}. The axisymmetric model cannot
rotate. It keeps its speed and hits the wall 32\% faster (0.264 against
0.200\,m/s), and the bounce is correspondingly too high (10.7 against
7.3\,mm). The restitution, which removes the effect of the impact velocity, is
much closer (0.71 against $0.68\pm0.05$). Scaled with $v_\mathrm{imp}^2$ to the
measured impact velocity, the simulated bounce would be 6.1\,mm. The maximal
indentation is $18.3\,\mu$m, above the Hertz value $13.3\,\mu$m of
table~\ref{tab:scales}, which is evaluated with the measured impact velocity.

\textbf{Rubber22, elastic.} The approach agrees to 1--3\% with experiment
and FSI. The indentation is 0.38\,mm, against the dry Hertz value
$\delta_H=0.33$\,mm; the difference is mainly the added mass of the fluid,
which the dry estimate does not contain. The lossless compliance returns
almost all of the stored energy: $e=0.85$, and the bounce is more than three
times too high. The model based on the static compliance of the ball is therefore not sufficient to reproduce the experimentally observed results. 
Material damping will be studied in the following section.

\begin{table}
\centering\small
\begin{tabular}{llccccc}
 & & $t_*$ [s] & $v_*$ [m/s] & $t_\mathrm{cont}$ [s] & $t_\mathrm{jump}$ [s] & $d_\mathrm{jump}$ [mm]\\
PTFE6 & experiment~\cite{WRFH2021} & 0.516 & $-0.331$ & 0.535 & 0.028 & 2.21\\
      & Eulerian FSI~\cite{WRFH2021} & 0.536 & $-0.316$ & 0.554 & 0.028 & 2.06\\
      & this model & 0.539 & $-0.314$ & 0.559 & 0.029 & 2.18\\
Rubber22 & experiment~\cite{WRFH2021} & 0.469 & $-0.309$ & 0.544 & 0.089 & 4.41\\
         & Eulerian FSI, $E=2$\,MPa & 0.450 & $-0.311$ & 0.522 & 0.091 & 5.50\\
         & Eulerian FSI, $E=5$\,MPa & 0.450 & $-0.311$ & 0.521 & 0.083 & 3.72\\
         & this model, $E=2$\,MPa, elastic & 0.456 & $-0.303$ & 0.531 & \PaperRubberSweepTjump & \PaperRubberSweepDjump\\
         & this model, $E=2$\,MPa, $\tan\delta=0.3$ & 0.456 & $-0.303$ & 0.530 & 0.099 & 7.54\\
         & this model, $E=2$\,MPa, $\tan\delta=0.5$ & 0.456 & $-0.303$ & 0.530 & 0.089 & 5.01\\
         & this model, $E=2$\,MPa, $\tan\delta=0.6$ & 0.456 & $-0.303$ & 0.530 & 0.085 & 4.01\\
\end{tabular}
\caption{Quantities of~\cite{WRFH2021}, Sec.~II~E, for the two cases studied
there. $d_\mathrm{jump}$ is the maximal film gap after contact. The damped runs
restart from the elastic run at patch activation (Section~\ref{sec:kvsweep}).}
\label{tab:fsi}
\end{table}

\subsection{Rubber22 with viscoelastic damping}\label{sec:kvsweep}

The elastic Rubber22 ball rebounds about three times too high. Real
materials, rubber in particular, dissipate part of the elastic energy by
internal friction, which we model by the Kelvin--Voigt compliance
\eqref{eq:kv} with $\tau>0$; the time discretisation is given in
Section~\ref{sec:splitting}. Material data usually give the loss factor
$\tan\delta$, i.e.\ the energy dissipated per load cycle divided by $2\pi$
times the maximal stored energy. For a harmonic load with frequency
$\omega$, $\tan\delta=\omega\tau$. For an impact of duration $t_c$ we take
$\omega\approx\pi/t_c$, i.e.\ $\tau\approx\tan\delta\;t_c/\pi$, with
$t_c\approx t_H$ from \eqref{eq:hertz}.
The energy balance
\eqref{eq:energy} gains the dissipation
$\tau\,\dot w\cdot\C_{cc}^{-1}\dot w\ge 0$, with
$\mathcal E_\mathrm{el}=\frac12 w\cdot\C_{cc}^{-1}w$. A Kelvin--Voigt solid
damps all wavelengths with the same $\tau$, so its dissipation grows with
frequency. A standard linear solid would limit this, at the cost of one
internal variable per patch node.

For Rubber22, $t_H=3.5$\,ms from~\eqref{eq:hertz}. The damping only acts
during contact, and the following simulations all start from the same
checkpoint at the first patch activation, shortly before the first contact.
Every damping case is continued at least until the second
contact.

\begin{table}
\centering\small
\begin{tabular}{ccccccc}
\hline
$\tan\delta$ & $\tau$ [ms] & $V_\mathrm{reb}/|V_\mathrm{imp}|$ & $d_\mathrm{jump}$ [mm] & $t_\mathrm{jump}$ [s] & $\Delta t_{12}$ [ms] & $h_{b,2}$ [mm]\\
\hline
0 & 0 & 0.81 & 14.10 & 0.132 & 315 & --\\
0.05 & 0.055 & 0.75 & 12.73 & 0.125 & 301 & --\\
0.1 & 0.111 & 0.71 & 11.29 & 0.118 & 281 & 2.36\\
0.2 & 0.222 & 0.62 & 9.17 & 0.106 & 250 & 1.44\\
0.3 & 0.332 & 0.55 & 7.51 & 0.099 & 226 & 0.94\\
0.5 & 0.555 & 0.44 & 4.97 & 0.088 & 185 & 0.44\\
0.6 & 0.666 & 0.40 & 3.98 & 0.084 & 170 & 0.29
\\
\hline
experiment & --- & $0.31\pm0.01$ & 4.41 & 0.089 & 175$^{a}$ & 0.22$^{a}$\\
Eulerian FSI, $E=2$\,MPa & --- & --- & 5.50 & 0.091 & --- & ---\\
Eulerian FSI, $E=5$\,MPa & --- & --- & 3.72 & 0.083 & --- & ---\\
\hline
\end{tabular}
\caption{Rubber22: sensitivity to material damping (Kelvin--Voigt), $p_\mathrm{
cav}=-1$\,bar, $r_c=0.493R$ (rule~\eqref{eq:rc}). The simulation-derived
rows are refreshed by \texttt{code/run\_paper\_sims.py}. $\Delta t_{12}$ is
the time between the first and second contact; $h_{b,2}$ is the second-bounce
height. Dashes in the comparison rows indicate quantities not reported in
those studies. $^{a}$~Run~1.}
\label{tab:kv}
\end{table}

The restitution decreases almost linearly with the loss factor, from 0.81
to 0.55 at $\tan\delta=0.3$ and 0.40 at $\tan\delta=0.6$, and the rebound
height follows. 

With $\tan\delta=0.6$ the rebound is almost reproduced: the first bounce is 4.3\,mm
against $4.0\pm0.2$\,mm (7\% high; $d_\mathrm{jump}=4.01$ against 4.41\,mm in
the quantities of~\cite{WRFH2021}). The time to the apex is 85 against
89\,ms. The second contact follows 170\,ms after the first, against 175\,ms
in the experiment, and the second bounce is 0.29 against 0.22\,mm. 

The loss factor $\tan\delta\approx0.6$ is the best fit to the experimental data. The damping of
this HNBR ball at the frequency of the impact ($\sim300$\,Hz) has not been
measured and a measurement would be required to validate the predictive character of the model for highly elastic material like rubber.


\subsection{A two-dimensional benchmark without experiment}\label{sec:bench2d}

As a second test case we consider a two-dimensional configuration published in~\cite{CFFFKSTW2026}. This setup is purely numerical and has no experimental counterpart. The authors have compared several numerical methods for the simulation of an elastic disc falling in a viscous fluid. Five groups with different models and discretization approaches participated in the benchmark. Table~\ref{tab:bench2d} cites the results of four of these groups which all lie within a narrow frame: 

[CF1] use a mixed Lagrangian-Eulerian approach with cut finite elements in the fluid, Nitsche coupling between fluid and solid and an augmented Lagrangian for imposing a minimum gap (comparable to the mesh size), implemented in FELiScE\footnote{Finite Elements for Life Sciences and Engineering \url{https://gitlab.inria.fr/felisce/felisce}}. 

[Fr1] uses a fully Eulerian approach, locally modified finite elements~\cite{J17,FreiRichter2014,Gozel2} for resolving the interface, Nitsche coupling and the same minimum gap, all implemented in Gascoigne 3d~\cite{gascoigne3d}. 

[KWF] is essentially the same approach, implemented in deal.II~\cite{dealII}. 

[FST] use an updated ALE approach with a remeshing strategy tuned for the contactless rebound without any further contact model~\cite{Fara2024}. Implementation is in FEniCS~\cite{fenics}. In every case we took the numbers corresponding to the finest discretization. No computing times are reported.

\textbf{Configuration.} A plane-strain elastic disk of radius
$r_s=0.011$\,m falls from rest in the box $(-0.04,0.04)\times(0,0.08)$\,m.
The bottom and side walls are no-slip and the top is free-slip. The centre
starts at height $h_0+r_s$ with $h_0=0.039$\,m. The fluid parameters are $\rho_f=1141$\,kg/m$^3$,
$\mu_f=0.008$\,Pa\,s and the solid's density is $\rho_s=1361$\,kg/m$^3$ with the Lam\'e parameters
$\mu_s=666.711$\,kPa and $\lambda_s=10$\,MPa, i.e.\ $E=1.958$\,MPa and
$\nu=0.469$. Gravity is $g=9.807$\,m/s$^2$. We compare with Model~I of \cite{CFFFKSTW2026} which is similar to the experimental benchmark. No-slip conditions act on both surfaces and we do not consider cavitation. The quantities of interest are those of
\cite[Table~1]{CFFFKSTW2026}. $t^*$, $v^*$ and $f^*$ are the time, the vertical
velocity and the vertical fluid force on the disk (including buoyancy) when the
gap equals $r_s$. $t_\mathrm{cont}$ and $d_\mathrm{cont}$ are the time and the size
of the minimal gap before the first rebound. $f_{\max}$ is the maximal
vertical force up to $t_\mathrm{cont}$, and $\max p_{bc}$ is the maximal pressure
at the centre of the bottom wall. $t_\mathrm{jump}$ and $d_\mathrm{jump}$ are the
time and the size of the maximal gap after $t_\mathrm{cont}$. Times are measured
from release.

With $f_{\max}\approx 65$\,N/m, the 2D Hertz estimate gives a contact half-width $a\approx 0.05\,r_s$ and an indentation of about $60\,\mu$m. Strains are a few percent, so linear elasticity is adequate and the patch model applies.

\textbf{Planar version of the model.} The implementation is a copy of the
axisymmetric one, with the radial weight and the hoop terms removed. We
compute the half domain $x>0$ with mirror symmetry at $x=0$ (no rotation, no
horizontal motion). The film is the planar Reynolds equation on
$0<x<r_c$. The patch radius follows from rule~\eqref{eq:rc} with the planar
Hertz pressure and the FE compliance below.

The one substantial change concerns the
compliance. In 3D the Boussinesq kernel decays like $1/r$, so the
half-space displacement vanishes at infinity. For $a\ll R$ this far-field
reference coincides with the centre of the ball up to $O(a/R)$, and the kernel
gives the approach of the body ($\delta=a^2/R$) from the patch alone. In 2D
the half-plane (Flamant) kernel is logarithmic~\cite[\S2]{Johnson1985},
\[
w(x)=-\frac{2}{\pi E^*}\int p(s)\ln|x-s|\dd s+\mathrm{const}.
\]
It grows
instead of decaying, so for a nonzero net force there is no reference at
infinity and the constant is undetermined. The kernel describes the shape of the
deformation but not its depth relative to the centre of mass, which we need for the coupled model. Therefore we replace $\C_{cc}$ by a finite element compliance of the actual free disk using the linear Navier--Lam\'e equations, discretized with $P_3$ elements on a mesh which is refined towards the lower contact surface (totalling about 78k dofs). $\C_{cc}$ is recomputed whenever the patch is activated. 
Every column of $\C_{cc}$ is computed by one solve: for column $j$, the film pressure $p=\phi_j(|x|)$ (the $P_1$ hat function of film node $j$) acts on the lower surface and is balanced by the uniform d’Alembert body force $-F_j/|\Omega_s|$, where $F_j$ is its net force. The normal displacement of this solution at the film nodes gives the column. The elasticity matrix is inverted only once and reused for all pressure nodes which makes the computation of $\C_{cc}$ very fast.

\begin{table}
\centering\small
\setlength{\tabcolsep}{3.2pt}
\begin{tabular}{lrrrrrrrrr}
 & $t^*$ & $v^*$ & $f^*$ & $t_\mathrm{cont}$ & $d_\mathrm{cont}$ & $f_{\max}$ & $\max p_{bc}$ & $t_\mathrm{jump}$ & $d_\mathrm{jump}$\\
coarse (3.5k dofs) & 0.28916 & $-0.17008$ & 4.9699 & 0.35630 & 1.27e-5 & 69.32 & 1.031e5 & 0.42174 & 1.024e-3\\
medium (7.9k)      & 0.28923 & $-0.17024$ & 4.9716 & 0.35625 & 1.27e-5 & 69.51 & 1.031e5 & 0.42089 & 1.012e-3\\
fine (21k)         & 0.28936 & $-0.17031$ & 4.9708 & 0.35637 & 1.26e-5 & 69.32 & 1.028e5 & 0.41955 & 0.968e-3\\
\hline
\multicolumn{8}{l}{\emph{Result of~\cite[Table~1]{CFFFKSTW2026}}:}\\
{[FST]}  & 0.28819 & $-0.17034$ & 4.9683 & 0.35736 & 5.92e-6 & 64.36 & 9.996e4 & 0.42883 & 1.321e-3\\
{[CF1]}  & 0.28925 & $-0.17000$ & 4.9681 & 0.35875 & 6.42e-6 & 64.07 & 9.667e4 & 0.42650 & 1.181e-3\\
{[KWF]}  & 0.28950 & $-0.17092$ & 4.9549 & 0.35801 & 1.15e-5 & 67.86 & 1.841e5 & 0.43375 & 1.520e-3\\
{[Fr1]}  & 0.28879 & $-0.17091$ & 4.9540 & 0.35777 & 4.05e-6 & 55.51 & 6.434e4 & 0.41776 & 0.931e-3\\
\end{tabular}
\caption{2D benchmark of~\cite{CFFFKSTW2026}: quantities of interest for the
full disk. Upper part: present model on three bulk meshes (degrees of freedom of the bulk). Lower part: finest levels of the no-slip (Model~I) approaches of
\cite[Table~1]{CFFFKSTW2026}; times in s, lengths in m, forces in N/m, pressures in Pa.}
\label{tab:bench2d}
\end{table}

\begin{figure}
\centering
\includegraphics[width=\textwidth]{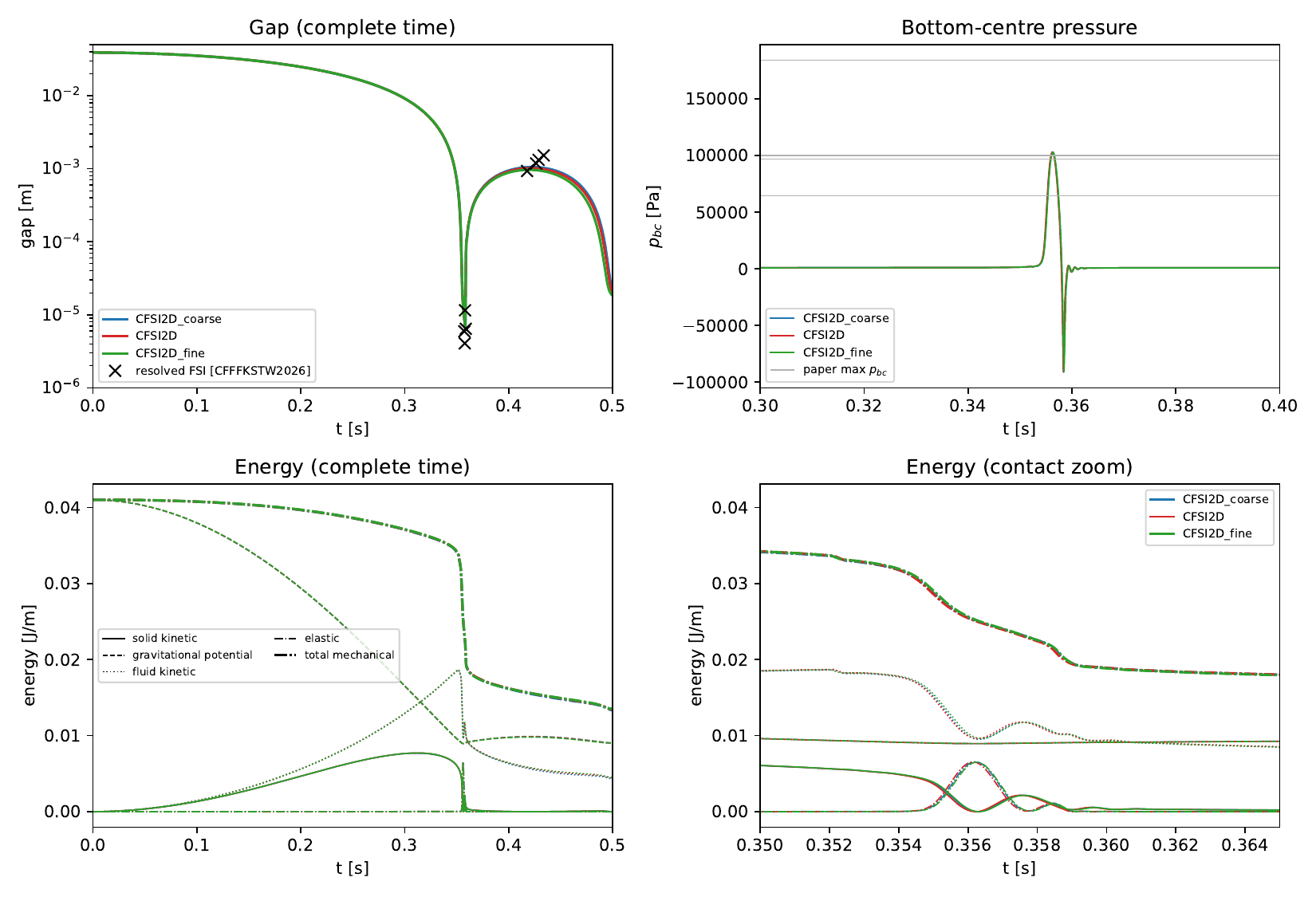}
\caption{2D benchmark on the three meshes. Top left: gap over the complete
fall; top right: pressure at the bottom centre from $0.30$ to $0.40$\,s. Bottom:
full-disk energy components and total mechanical energy over the complete
fall (left) and from $0.35$ to $0.365$\,s (right). The total is the sum of solid kinetic,
gravitational potential, fluid kinetic and elastic energies. Crosses mark
$(t_\mathrm{cont},d_\mathrm{cont})$ and $(t_\mathrm{jump},d_\mathrm{jump})$ from the
approaches in table~\ref{tab:bench2d}; horizontal lines show the paper's
maximum bottom-centre pressures.}
\label{fig:bench2d}
\end{figure}

\textbf{Results.} Table~\ref{tab:bench2d} and figure~\ref{fig:bench2d}
show the results. On the three meshes nearly all quantities lie within the ranges reported in~\cite{CFFFKSTW2026}. 

The peak force $f_\mathrm{max}$ gives a slightly larger value, 8--9\% for [FST] and [CF1], about 2\% for [KWF]. [Fr1] shows a significant deviation from the range of the other approaches. The rebound height $d_\mathrm{jump}$ is at the lower end of the range, but still within it. The best agreement is with [Fr1], which corresponds to the good agreement with the computations of Frei in the PTFE6 case above. 

Computational times are reported in table~\ref{tab:runtime}. The fine mesh takes about 30 minutes, the medium mesh about 3.5 minutes and the coarse mesh takes 1.3 minutes, still giving results close to the resolved simulation. 

The bottom row of figure~\ref{fig:bench2d} shows the energy evolution and agrees well with~\cite[figures 5 \& 5]{CFFFKSTW2026}. The total energy is non-increasing, except for local increases of at most $1.2\cdot10^{-3}E(0)$ per step on the coarse mesh and
  $3\cdot10^{-4}E(0)$ on the medium and fine meshes, at the end of the contact, mostly due to the switch-off of the patch. 
Without any adjustment of parameters, the reduced model thus reproduces the
resolved benchmark.

\begin{table}
\centering\small
\begin{tabular}{lccc}
\hline
simulation & wall time [min] & time steps & mean bulk DOFs\\
\hline
PTFE6 (axisymmetric) & 9.1 & 1760 & 9\,019\\
PTFE10 (axisymmetric) & 5.5 & 1213 & 7\,379\\
POM20 (axisymmetric) & 10.0 & 1920 & 9\,546\\
Rubber22 elastic (axisymmetric) & 7.6 & 1672 & 8\,033\\
2D benchmark, coarse & 1.4 & 876 & 3\,885\\
2D benchmark, medium & 3.6 & 1113 & 8\,590\\
2D benchmark, fine & 23.7 & 1820 & 23\,200
\\
\hline
\end{tabular}
\caption{Wall-clock times for the full, independent simulations, measured on a single core of a MacBook Pro M5 (2025). The \emph{mean bulk DOFs} are
averaged over the time steps and count the degrees of freedom of the bulk
fluid discretization only (without the Reynolds film unknowns and, in 2D,
without the precomputed compliance of the elastic disk).}
\label{tab:runtime}
\end{table}

\section{Discussion and outlook, Limits of the model}\label{sec:discussion}\label{sec:limits}

The model is intentionally light-weight and should be able to scale to full 3D simulations including the contact of multiple bodies. It uses a rigid ball with a fixed half-space compliance on the patch, an axisymmetric bulk flow and smooth surfaces. A run needs about 10k degrees of freedom and takes minutes on one core. Linear systems are currently solved using a direct solver requiring re-decomposition in every time step due to the nonlinearity. Efficient (parallel) implementation based on iterative solvers will allow for upscaling to many bodies and real 3D (including rotation). 

A rigid body with an asymptotic film and a
fixed compliance reproduces contact-free rebound with the correct
magnitude for a stiff ball. For PTFE6 the final code is within 1\% of the
Eulerian FSI before contact, and the rebound height agrees with the
experiment within the scatter of the runs. For PTFE10 the approach agrees
to 1--2\% and the rebound is 12\% too high. There is no contact parameter. Even the patch radius is fixed a priori by rule~\eqref{eq:rc}, the same for all
configurations in 3D and 2D, and the results depend only weakly on it
above a sharp lower limit that the code detects
(Section~\ref{sec:rctest}). 

\textbf{Rotation and three-dimensional motion are missing.} The POM ball
  rotates by about $180^\circ$ during settling and decelerates strongly in
  front of the wall~\cite{HTR2021}. The axisymmetric model cannot represent
  this. It hits the wall 32\% too fast, and the bounce is 46\% too high. This
  case needs a 3D model with rotation. Including rotational degrees of freedom will make the numerical model more challenging as the elimination of the patch pressure will no longer be possible by means of a $2\times 2$ system.

\textbf{The model loses too little energy in the contact.} The
  restitution is too high for all four balls: 0.50 against 0.45 (PTFE6), 0.71
  against 0.63 (PTFE10), 0.71 against 0.68 (POM20), even where the approach is
  correct. For Rubber22, we observe 0.81 against 0.31 in the case of static elasticity and 0.40 using the Kelvin--Voigt model with $\tan\delta=0.6$. Two effects which could be relevant are not in the model:
\emph{1. Surface roughness.} The minimal film thickness of the
    smooth-surface model (0.27--0.36\,$\mu$m for PTFE and POM, 4.7\,$\mu$m for
    rubber) is below the roughness of the balls (0.87--0.88\,$\mu$m and
    3.84\,$\mu$m~\cite{HTR2021}). It must therefore be expected that contact will take place in the experiments, with an effect on the flow in the film, dissipating energy. The Reynolds film model is formally not valid. 
  \emph{2. Material damping.} The compliance is lossless. For the
    polymers the loss will be small but for the rubber ball it can be a dominant effect.

  \textbf{Soft bodies exceed the linear half-space.} For Rubber22 the
  flattening reaches $a_H/R=0.17$, and the indentation is 0.38\,mm. The fixed
  Boussinesq compliance on the patch then represents neither the finite size
  of the body nor its near-incompressibility exactly. Without damping the
  rebound is more than three times too high. With a fitted loss factor
  ($\tan\delta\approx0.6$) height and timing of both bounces agree, but this
  fit also absorbs the other errors of the soft case.

These model weaknesses can be attributed to missing effects and not to the splitting between bulk flow and surface interactions. Rotation in 3D, a roughness model in the film and a viscoelastic compliance can be added without changing the general structure of the method. At the present level the model already predicts the
settling, the time of impact and the order of magnitude of the rebound for
stiff balls. The computational time is a fraction of that of fully resolved elastic fluid-structure interaction simulations. 

Several open points remain for future work: 
An important limitation of the present model is the restriction to axisymmetric motion. The POM ball rotates by about $180^\circ$ during settling and decelerates strongly in front of the wall~\cite{HTR2021}. The axisymmetric model cannot represent this. It hits the wall 32\% too fast, and the bounce is 46\% too high. This case needs a 3D model with rotation. Another apparent model deficiency is the assumption of surface smoothness, which is not the case for the PTFE and POM balls: surface roughness is of the order of the minimal film thickness, so contact will take place in the experiments, with an effect on the flow in the film, dissipating energy. 

In terms of the numerical discretization, the splitting is first order in time and does not satisfy the geometric conservation law (GCL). The GCL is important for long-time simulations of many bodies. 

Several components like the cavitation model or the reinitialization after deactivating the patch are first simple realizations. They can be improved, e.g.\ by a mass-conserving cavitation model or an adaptive patch radius during separation for soft bodies. 


\section*{Acknowledgements}

The work was supported by the Deutsche Forschungsgemeinschaft (DFG, German Research Foundation) under grant number 537063406. 

\section*{Data availability}

The implementation for reproducing the numerical results is openly available in the Zenode repository~\cite{implementation}. This repository contains a description of the implementation and scripts for running all experiments presented in this manuscript.

\section*{Use of AI}

Claude Opus 5.5 and Claude Code have been used to assist in writing, documenting and maintaining the numerical implementation documented in~\cite{implementation}. The author has reviewed and edited the source code and takes full responsibility for the content of the code and the scientific results presented in this paper. Furthermore, Claude Opus 5.5 has been used for improving the English language of the manuscript. 

\section*{Declaration of interests}

The authors report no conflict of interest.

\appendix

\section{Sensitivity to the parameters}\label{app:sensitivity}

\subsection{Sensitivity to the patch radius}\label{sec:rctest}

Apart from the switching gaps (Section~\ref{sec:switchtest}), the patch
radius is the only numerical parameter of the splitting that is not a
discretisation parameter. Rule~\eqref{eq:rc} fixes it through one
constant $C_{rc}$. To test the rule we vary $C_{rc}$ between 2 and 8 for
two cases: PTFE10 (axisymmetric, Boussinesq compliance, St $\approx130$) and
the 2D benchmark (planar, FE compliance, medium mesh). 

\begin{table}
\centering\small
\setlength{\tabcolsep}{4pt}
\begin{tabular}{ccccccc}
\multicolumn{7}{l}{\emph{PTFE10 (3D axisymmetric)}}\\
$C_{rc}$ & $r_c/R$ & $e$ (1\,kHz) & $V_\mathrm{reb}/|V_\mathrm{imp}|$ & $h_b$ [mm] & $t_\mathrm{jump}$ [ms] & $h_{\min}$ [$\mu$m]\\
\hline
2 & 0.142 & \multicolumn{5}{c}{bulk layer at $\Sigma$ closes (guard)}\\
3 & 0.174 & 0.716 & 0.774 & 7.93 & 52.3 & 0.363\\
4 & 0.201 & 0.714 & 0.776 & 7.85 & 52.3 & 0.362\\
6 & 0.246 & 0.716 & 0.777 & 7.95 & 53.4 & 0.363\\
8 & 0.285 & 0.714 & 0.777 & 7.94 & 53.4 & 0.364\\
\hline
\multicolumn{7}{l}{\emph{2D benchmark (planar, medium mesh)}}\\
$C_{rc}$ & $r_c/r_s$ & $d_\mathrm{cont}$ [m] & $f_{\max}$ [N/m] & $\max p_{bc}$ [Pa] & $t_\mathrm{jump}$ [s] & $d_\mathrm{jump}$ [m]\\
\hline
2 & 0.188 & \multicolumn{5}{c}{bulk layer at $\Sigma$ closes (guard)}\\
3 & 0.230 & 1.27e-5 & 68.76 & 1.033e5 & 0.42149 & 1.031e-3\\
4 & 0.266 & 1.27e-5 & 69.51 & 1.031e5 & 0.42089 & 1.012e-3\\
6 & 0.325 & 1.26e-5 & 70.01 & 1.024e5 & 0.41904 & 0.957e-3\\
8 & 0.376 & 1.29e-5 & 70.27 & 1.017e5 & 0.41786 & 0.919e-3\\
\end{tabular}
\caption{Sensitivity to the constant $C_{rc}$ of rule~\eqref{eq:rc}
($C_{rc}=4$ is the default used everywhere else). $r_c$ is the patch radius
at the first contact.}
\label{tab:rctest}
\end{table}

The results are summarised in table~\ref{tab:rctest}.
\emph{Lower limit.} For $C_{rc}=2$ the run stops in both cases: the
  elastic flattening reaches the patch edge and the bulk layer at $\Sigma$
  closes. The lower limit is thus the same, $C_{rc}\approx2$--$3$, in 3D and
  in 2D, although the patch radius differs ($0.14R$ against $0.19r_s$) and
  although the 2D approach is several times larger than $a^2/R$. 
\emph{PTFE10.} Above the limit the results are practically independent
  of the patch radius. Rebound height and restitution vary by at most 1.3\%,
  the minimal film thickness by 0.5\%.
\emph{2D.} Approach velocity, minimal gap, peak force and peak pressure vary by
  about 2\%. The rebound height decreases monotonically with the patch
  radius, by 11\% between $C_{rc}=3$ and $8$. In 2D the film is thicker at the
  patch edge for a given $r_c$, and in a larger patch a larger part of the
  separating flow is described by the inertia-free film. 

The patch radius is therefore not a classical tuning parameter.
Below a sharp lower limit the model is inconsistent, and the simulation crashes. Above it the results depend only weakly on $r_c$, with the rebound in the planar case as the most sensitive quantity. $C_{rc}=4$ is a safe
distance from the limit, and the same value is used for all configurations.

\subsection{Sensitivity to the switching gaps}\label{sec:switchtest}

The switching rule \eqref{eq:dact}, \eqref{eq:switch_off} contains the
constants $d_\mathrm{act}$, $\varepsilon_\mathrm{act}$, $d_\mathrm{deact}$ and the
factor $0.75$. We vary the switch-on gap $d_\mathrm{act}^\mathrm{eff}$ with the
switch-off gap fixed, and the switch-off gap $d_\mathrm{deact}^\mathrm{eff}$ with
the switch-on gap fixed, for the 2D benchmark on the medium mesh. All variants
and the reference case end at $0.45$\,s, shortly after the first rebound.
With the default values the
patch is switched on by the elasticity criterion at
$d_\mathrm{act}^\mathrm{eff}=0.038\,r_s$ and switched off at
$d_\mathrm{deact}^\mathrm{eff}=0.75\,d_\mathrm{act}^\mathrm{eff}=0.029\,r_s$, where
$h(r_c)/r_c\approx0.24$ ($r_c=0.266\,r_s$).

\begin{table}
\centering\small
\setlength{\tabcolsep}{3pt}
\begin{tabular}{lcccccc}
\toprule
variant & $d^\mathrm{eff}/r_s$ & $d_\mathrm{cont}$ [m] & $f_{\max}$ [N/m] & $\max p_{bc}$ [Pa] & $t_\mathrm{jump}$ [s] & $d_\mathrm{jump}$ [mm]\\
\hline
default & 0.038 / 0.029 & $1.27\cdot10^{-5}$ & 69.5 & $1.03\cdot10^{5}$ & 0.4209 & 1.009 \\
\hline
\multicolumn{7}{l}{\emph{switch-on gap varied, $d_{\rm deact}^{\rm eff}=0.029\,r_s$}}\\
$\varepsilon_{\rm act}=3\cdot10^{-4}$ & 0.063 & $-0.3$ & $+0.6$ & $+0.3$ & $+0.1$ & $+2.4$ \\
$\varepsilon_{\rm act}=3\cdot10^{-3}$ & 0.024 & $+1.0$ & $-0.6$ & $-0.2$ & $-0.1$ & $-1.3$ \\
$\varepsilon_{\rm act}=10^{-2}$ & 0.015 & $+0.6$ & $-1.7$ & $-0.8$ & $-0.5$ & $-6.9$ \\
$\varepsilon_{\rm act}=0$, $d_{\rm act}=0.01\,r_s$ & 0.010 & $-1.0$ & $-1.4$ & $-0.6$ & $-0.3$ & $-5.4$ \\
\hline
\multicolumn{7}{l}{\emph{switch-off gap varied, $d_{\rm act}^{\rm eff}=0.038\,r_s$}; $h(r_c)/r_c$ at switch-off in brackets}\\
$d_{\rm deact}^{\rm eff}=0.010\,r_s$ & (0.17) & $+0.1$ & $+0.1$ & $+0.1$ & $+3.0$ & $+45.5$ \\
$d_{\rm deact}^{\rm eff}=0.015\,r_s$ & (0.19) & $+0.0$ & $+0.0$ & $+0.0$ & $+0.5$ & $+5.6$ \\
$d_{\rm deact}^{\rm eff}=0.020\,r_s$ & (0.21) & $-0.1$ & $+0.1$ & $+0.1$ & $+0.2$ & $+1.7$ \\
$d_{\rm deact}^{\rm eff}=0.040\,r_s$ & (0.29) & $+0.0$ & $+0.1$ & $+0.1$ & $-0.2$ & $-0.8$ \\
$d_{\rm deact}^{\rm eff}=0.060\,r_s$ & (0.36) & $+0.0$ & $-0.1$ & $+0.0$ & $-0.3$ & $-2.8$ \\
\end{tabular}
\caption{2D benchmark: sensitivity to the switching gaps on the medium mesh.
The default row gives absolute quantities; the variants give changes in \%
relative to that default. For switch-off rows, $h(r_c)/r_c$ at deactivation is
shown in parentheses. The generated rows are refreshed by
\texttt{code/run\_paper\_sims.py}.}
\label{tab:switchtest}
\end{table}

\emph{Switch-on.} The approach and the impact quantities depend only weakly
on the switch-on gap: over a factor of six in $d_\mathrm{act}^\mathrm{eff}$ the
peak force and pressure change by at most 2\%. The rebound height
increases monotonically with the switch-on gap, by about 10\% between
$0.010$ and $0.063\,r_s$. For $\varepsilon_\mathrm{act}=3\cdot10^{-4}$ the
integration ends at $0.45$\,s, shortly after the first rebound. Continuing
into the second approach triggers the patch-edge guard of
Section~\ref{sec:impl}, which also responds to rigid approach, not only
elastic flattening. All quantities reported here are obtained before this
cutoff.

\emph{Switch-off.} Above $d_\mathrm{deact}^\mathrm{eff}\approx0.02\,r_s$ the
rebound is practically independent of the switch-off gap (within 2.5\%),
up to $h(r_c)/r_c=0.36$; in 2D no instability of the film--bulk coupling
occurred in this range. Below, the rebound increases sharply and
independently of the mesh, by 44\% for $0.010\,r_s$. The ball velocity is
continuous at the switch in all runs ($\le0.3\%$), but the fluid kinetic
energy is not: the re-added gap region is filled with the squeeze flow
($u_x\sim V x/h$), whose kinetic energy the inertia-free film does not
carry. At $0.010\,r_s$ this jump amounts to 8\% of the kinetic energy at
impact, ten times more than for the default, and part of it reappears in
the rebound. The switch-off gap therefore has a lower bound as well as the
upper bound $h(r_c)/r_c\ll1$: the kinetic energy of the restored gap flow
must be small compared with the kinetic energy of the ball.

In summary, within $0.015\le d_\mathrm{act}^\mathrm{eff}/r_s\le0.063$ and
$0.02\le d_\mathrm{deact}^\mathrm{eff}/r_s\le0.06$ all quantities except the
rebound change by at most 2\%, and the rebound height by at most 7.5\%. This
is of the same size as the dependence on the patch radius and well inside
the spread of the reference results (0.93--1.52\,mm). The default switch-off
gap lies inside the plateau; the purely geometric value
$d_\mathrm{deact}=0.02R$ is at its lower end.

\bibliographystyle{plain}

\end{document}